\documentclass[preprint]{aastex}

\input{psfig.sty}
\def\etal   {{\it et~al.\/}}

\def\HII    {H~{\rm {II}}}
\def\kms    {~km~s$^{-1}$}
\def\mo     {{$M_{\odot}$}}

\begin{document}

\title{The DEEP Groth Strip Survey XII: 
The Metallicity of Field Galaxies at $0.26<z<0.82$
and the Evolution of the Luminosity-Metallicity Relation}

\author{Henry A. Kobulnicky}
\affil{Department of Physics \& Astronomy \\ 
University of Wyoming \\ Laramie, WY 82071 
\\ Electronic Mail: chipk@uwyo.edu}

\author{Christopher  N.~A. Willmer\footnote{On 
leave from Observat\'orio Nacional, Rio de Janeiro, Brazil}, 
Benjamin J. Weiner, \\ 
David C. Koo, Andrew C. Phillips, \& S.~M. Faber }
\affil{University of California, Santa Cruz \\ Department of Astronomy \& Astrophysics \\ 
Santa Cruz, CA, 95064  \\ 
cnaw@ucolick.org, bjw@ucolick.org, koo@ucolick.org,  
phillips@ucolick.org, faber@ucolick.org}

\author{Vicki L. Sarajedini}
\affil{Department of Astronomy \\ University of Florida \\
Gainesville, FL 32611 \\  vicki@astro.ufl.edu}

\author{Luc Simard}
\affil{Canadian Astronomy Data Centre \\
Herzberg Institute of Astrophysics \\ National Research Council of Canada  \\ 
 luc.simard@nrc.ca }

\author{Nicole P. Vogt} 
\affil{Department of Astronomy \\ New Mexico State University \\
	P.O. Box 30001, Dept 4500 \\  Las Cruces, NM 88003-8001 \\ 
nicole@nmsu.edu}

\author{Revised draft of 28 April 2003}



\vskip 1.cm

\begin{abstract}

Using spectroscopic data from the Deep Extragalactic Evolutionary
Probe (DEEP) Groth Strip survey (DGSS), we analyze the gas-phase
oxygen abundances for 56 emission-line field galaxies in the redshift
range $0.26<z<0.82$.  These galaxies comprise a small subset selected
from among 693 objects in the DGSS.  They are chosen for chemical
analysis because they exhibit the strongest emission lines, and thus
have the highest star formation rates per unit luminosity. Oxygen
abundances relative to hydrogen range between $8.4<12+log(O/H)<9.0$
with typical uncertainties of 0.17 dex.  The 56 DGSS galaxies
collectively exhibit a correlation between B-band luminosity and
metallicity, i.e., an L-Z relation.  Using the DGSS sample and local
galaxy samples for comparison, we searched for a ``second parameter''
which might explain some of the dispersion in the L-Z relation.
Parameters such as galaxy color, emission line equivalent width, and
effective radius were explored but found to be uncorrelated with
residuals from the the mean L-Z relation.  The source of the
dispersion in the L-Z relation is either intrinsic or depends upon a
yet-unidentified combination of parameters.

Subsets of DGSS galaxies binned by redshift also exhibit L-Z
correlations but with different zero points.  Galaxies in the highest
redshift bin ($z=0.6-0.82$) are brighter by $\sim1$ mag compared to
the lowest redshift bin ($z=0.26-0.40$) and brighter by $\sim1-2$ mag
compared to local ($z<0.1$) field galaxies.  This offset from the
local L-Z relation is greatest for objects at the low-luminosity
($M_B\sim-19$) end of the sample, and vanishingly small for objects at
the high-luminosity end of the sample ($M_B\sim-22$).  Thus, both the
slope and zero point of the L-Z relation appear to evolve.  Based on
the DGSS galaxies, we provide an approximate analytic expression for
the mean gas-phase metallicity of galaxies as a function of luminosity
and redshift---a recipe that may aid in modeling the properties of
cosmologically distant galaxies.  Emission-line field galaxies have
undergone moderate amounts of evolution in the past $\sim7$ Gyr since
z$\sim0.8$, and the evolution is most significant for galaxies of
lower luminosity. Either the least-luminous DGSS field galaxies have
faded by 1--2 mag due to decreasing levels of star formation, or they
have experienced an increase in the mean metallicity of the
interstellar medium by factors of 1.3--2 (0.1-0.3 dex).  Plausible
evolutionary models suggest that a combination of the two processes is
likely.  The relatively greater degree of luminosity and metallicity
evolution seen among the lower luminosity (sub $L*$) galaxies in the
last 8 Gyr implies either a more protracted assembly process, or a
more recent formation epoch compared to more luminous $L*$ galaxies.

\end{abstract}


\section{Metallicity as a Measure of Galaxy Evolution} 

Many recent research programs in galaxy evolution trace changes in
correlations between fundamental galaxy properties as a function of
cosmic epoch.  One such approach compares the number density and
luminosity function of galaxies at earlier times (Lilly
\etal\ 1995; Bershady \etal\ 1997; Sawicki, Lin, \& Yee 1997; Lin
\etal\ 1999; ) to the luminosity function in the nearby universe (Zucca
\etal\ 1997; Marzke \etal\ 1998; Norberg \etal\ 2002).  Evidence suggests 
an increase in the number density of small, blue galaxies at earlier
times with only a small amount of passive fading among the more
luminous, redder galaxies.  Other investigations compare the relation
between rotation velocity and luminosity (T-F relation; Tully \&
Fisher 1977) locally with that observed in more distant disk galaxies
(e.g., Forbes \etal\ 1995; Vogt \etal\ 1996, 1997; Simard \& Pritchet 1998).
Most results indicate that galaxies of a given rotational amplitude
appear 0.2-1.0 mag brighter at $z\sim1$, although Bershady \etal\
(1998) argue that the variation is significant for only the bluest
galaxies, and Kannappan, Fabricant \& Franx (2002) show that
Tully-Fisher residuals are strongly correlated with galaxy color.
Simard \etal\ (1999) find no evidence for evolution of the
size-magnitude relation among 190 field galaxies out to $z=1.1$ in the
Groth Strip (Groth 1994).  They show that selection effects favor high
surface brightness galaxies in existing surveys.  These biases may
mimic the appearance of luminosity evolution and/or surface brightness
evolution reported in many studies (Schade \etal\ 1996a, 1996b; Roche
\etal\ 1998; Lilly
\etal\ 1998).  Yet another line of investigation compares the
fundamental plane\footnote{The fundamental plane is the locus
populated by early type galaxies galaxies in the three-dimensional
parameter space of surface brightness, velocity dispersion, and size
(parameterized by effective radius) (Dressler \etal\ 1987; Djorgovski
\& Davis 1987).} correlations for early-type galaxies as a function of
epoch.  Recent results suggest an decrease in the mean mass-to-light
ratio with lookback time for elliptical galaxies (van Dokkum \& Franx
1996; Kelson \etal\ 1997).  In Paper X of this series, Im \etal\
(2001) found a brightening of 1-2 mag for early-type $L^*$ galaxies
out to $z\sim1$.

In this paper we focus on the evolution of the correlation between
metallicity, $Z$, and total luminosity, i.e., the
luminosity-metallicity (L-Z) relation to which nearly all types of {\it
local} galaxies conform.  In early-type systems, absorption line indices
provide a measure of the mean stellar iron or magnesium abundances
(Faber 1973; Brodie \& Huchra 1991; Trager 1999) while in spiral
(Zaritsky, Kennicutt, \& Huchra 1994) and dwarf systems (Lequeux
\etal\ 1979; French 1980; Skillman, Kennicutt, \& Hodge 1989; Richer \&
McCall 1995) the \HII\ region oxygen abundance relative to hydrogen is
the basis of metallicity measurement.  The luminosity-metallicity
relation (L-Z) can be used as a sensitive probe and consistency check 
of galaxy evolution.
The metallicity of a galaxy can only increase monotonically with time
(unless large-scale infall of primordial gas is invoked), while the
luminosity may increase or decrease depending on the instantaneous star
formation rate.  Metallicity is less sensitive to variations due
to transient star formation events in a galaxy's history.  Most models and
observations of galaxies at earlier epochs predict higher star
formation rates and a larger fraction of blue, star-forming galaxies in
the past (Madau \etal\ 1996; Lilly \etal\ 1996; Somerville \& Primack
1999).  Thus, a high or intermediate-redshift galaxy sample ought to
be systematically displaced from the local sample in the
luminosity-metallicity plane if individual galaxies participate in
these cosmic evolution processes.   However, if local effects such as
the gravitational potential and ``feedback'' from winds and supernovae
regulate the star formation and chemical enrichment process, then the
L-Z relation might be independent of cosmic epoch.  
Semi-analytic models of Somerville \& Primack (1999), for example, show
very little evolution of the L-Z relation with redshift.
One goal of this paper is to provide data capable of 
testing these alternative hypotheses.

At high redshift, absorption line surveys have been tracing the
chemical evolution of the Lyman-alpha forest and damped Lyman-alpha
systems for more than a decade (Bergeron \& Stasi\'nska 1986; Sargent
\etal\ 1988; Steidel 1990; Pettini \etal\ 1997; Lu, Sargent, \& Barlow
1997; Prochaska \& Wolf 1999), while several studies have recently begun to
trace the cosmic chemical evolution of {\it individual galaxies} with
known morphological and photometric properties.  Kobulnicky \& Zaritsky
(1999) measured \HII\ region oxygen and nitrogen abundances for a sample
of 14 compact starforming galaxies with kinematically narrow emission
lines in the range $z=0.1-0.5$.  Their sample was found to conform to
the local L-Z relation, within observational uncertainties.  Carollo \&
Lilly (2001) studied 13 star-forming galaxies at $0.5<z<1$ from the
Canada-France Redshift Survey (CFRS; Lilly \etal\ 1995) and found no
significant evidence for evolution in the L-Z relationship out to z=1.
However, at larger redshifts of $z>2$, both Kobulnicky \& Koo (2000)
and Pettini \etal\ (2001) found that Lyman break galaxies with
metallicities between $0.1~Z_\odot<Z<0.8~Z_\odot$ are 2-4 magnitudes
more luminous than local galaxies of similar metallicity.  This
deviation from the local L-Z relation demonstrates that the
luminosity-to-metal ratio varies throughout a galaxy's lifetime and is a
potentially powerful diagnostic of its evolutionary state.

The DEEP (Deep Extragalactic Evolutionary Probe; Vogt \etal\ 2003;
Paper I) team has been assembling Keck spectroscopic data on galaxies
in the Groth Strip survey (Groth 1994) in order to study the evolution
of field galaxies.  Previous papers in this series include a study of
the evolution of the Fundamental Plane (FP) of early-type galaxies
(Paper IX; Gebhardt \etal\ 2003), the evolution of
the luminosity function of E/S0 galaxies (Paper X; Im \etal\ 2001), and
the structural parameters of Groth Strip galaxies (Simard \etal\ 2002;
Paper II).  In this paper, we explore the chemical properties of 56
star-forming emission-line field galaxies observed as part of the DEEP
Groth Strip Survey (DGSS).  We measure the interstellar medium oxygen
abundances from the [O~II], [O~III], and H$\beta$ emission lines to
analyze the degree of metal enrichment as a function of redshift,
luminosity, and other fundamental parameters.  Throughout, we adopt a
cosmology with $H_0$=70 \kms\ Mpc$^{-1}$, $\Omega_m=0.3$,  and
$\Omega_\Lambda=0.7$.

\subsection{Target Selection and Observations}

The DEEP Groth Strip Survey (DGSS) consists of Keck spectroscopy over
the wavelength range $\sim$4400 \AA\ -- 9500 \AA\ obtained with the Low
Resolution Imaging Spectrometer (LRIS; Oke \etal\ 1995).  The spectra have
typical resolutions of 3-4 \AA.  Integration times ranged
from 3000 s to 18,000 s with a mean of around 6000 s.   A full
description appears in Vogt \etal\ (Paper I in this series; 2003).

We searched the spectroscopic database for galaxies in the Groth Strip
Survey with nebular emission lines suitable for chemical analysis.
Only galaxies where it was possible to measure all of the requisite
[O~II]$\lambda3727$, H$\beta$, [O~III]$\lambda$4959, and
[O~III]$\lambda$5007 lines were retained.   These criteria
necessarily exclude objects at redshifts of $z\lesssim0.26$ since the
requisite [O~II]$\lambda$3727 line falls below the blue limit of the
spectroscopic setup.  Likewise, objects with redshifts $z\gtrsim 0.82$ are
excluded because the [O~III]$\lambda$5007 line falls beyond the red
wavelength limit of the survey.  As of February 2002, there were 693
objects with Keck spectra with identified redshifts in the DEEP Groth
Strip Survey.  Of these 693 objects, 398 unique objects have
spectroscopic redshifts between within the nominal limits
$0.26<z<0.82$.  The usable sample is further reduced because 23
candidates were positioned on the slitmask such that the [O~III] lines
fell off the red end of the spectral coverage.  An additional 49
objects were rejected because their position on the slitmask caused the
[O~II]$\lambda$3727 line to fall off the blue end of the spectral
coverage.  Furthermore, atmospheric $O_2$ absorption troughs between
6865 \AA\ -- 6920 \AA\ (the ``B band'') and between 7585 \AA\ --
$\sim7680$ \AA\ (the ``A band'') prohibit accurate measurement of
emission lines for objects in particular redshifts ranges.  We removed
14 objects from the sample in the redshift interval $0.410<z<0.426$
where the $H\beta$ line falls in the B band.  We removed 8 additional
objects in the redshift interval $0.56<z<0.58$ where $H\beta$ 
falls in the A band.  We removed 24 objects in the redshift interval
$0.52<z<0.54$ which places both [O~III] $\lambda$4959 and [O~III]
$\lambda$5007 in the A band.  The B band is sufficiently narrow that
either [O~III] $\lambda$4959 or [O~III] $\lambda$5007 is always
available for measurement.  Following this selection process, 276
objects remain.

A further list of 210 objects were removed from the sample because the
$H\beta$ emission line was absent or too weak (S/N$<$8:1) for reliable
chemical determinations (Kobulnicky, Kennicutt, \& Pizagno 1999 for a
discussion of errors and uncertainties).   The spectra of objects
rejected due to a weak $H\beta$ line are usually dominated by stellar
continuum rather than nebular emission from star-forming regions.  Most
local early-type spirals and elliptical galaxies share these spectral
characteristics.  For these objects, H$\beta$ is seen in absorption
against the stellar spectrum of the galaxy.  Thus early type galaxies with
older stellar populations are preferentially rejected in favor of late
type galaxies with larger star formation rates.  There should, however,
be no metallicity bias introduced by rejecting these 210 objects since
we will compare the DGSS sample with a local sample selected in the
same manner: on the basis of emission lines measured with a signal-to-noise 
of 8:1 or better.

Objects with strong $H\beta$ but immeasurably weak [O~II] or [O~III]
lines present a data selection conundrum.  In principle, such objects
should be included in the sample to avoid introducing a metallicity
bias, but it is not possible to compute metallicities if the oxygen
lines are not detected.  Only 3 such objects were found in
the database.  For these objects, it appears that poor sky subtraction
caused the [O~III] features to be immeasurably weak.
Intrinsically-weak oxygen lines may be caused by either extremely high
($Z>2~Z_\odot$) or extremely low ($Z<0.05~Z_\odot$) metal content.  In
the latter case, oxygen lines are weak because of the lack of $O^{+}$
and $O^{++}$ ions.  However, in the local universe,
galaxies with extremely low intrinsic
abundances are under-luminous ($M_B>-15$), and no such faint galaxies
are included in our sample.  In the high-metallicity case, efficient
cooling decreases the mean collisional excitation level, reducing the
[O~III] line strengths.  However, the Balmer lines and [O~II] lines
are not strongly affected by reductions in electron
temperature.  It is unlikely that our rejection criteria bias the
sample by preferentially excluding either metal-poor or metal-rich
systems.

Of the original 693 objects, 66 galaxies remain.  These objects appear
in Table~1, along with their equatorial coordinates, redshift,
absolute blue magnitude, restframe $(B-V)_0$ color, half-light radius
$R_{hl}$, and bulge fraction $B/T$ derived from model fitting routines
(Simard \etal\ 1999, 2002).  The restframe ($B--V)_0$ and ($B--R)_0$ colors
used in this work were calculated following the procedure of Lilly
\etal\ (1995), by interpolating the measured colors of DGSS galaxies
over a subset of Kinney \etal\ (1996) spectra, which were then used to
synthesize the different restframe colors.  Figure~1 shows their
redshifted spectra with major emission lines identified next to the
F814W HST greyscale images.  A cursory glance at the images reveals an
assortment of galaxy types, from small compact objects to spiral disks
and obvious mergers-in-progress.

In order to assess whether the 66 selected objects are representative
of the 398 galaxies with spectra in the $0.25<0.82$ redshift range,
Figure~\ref{hist} shows histograms of their morphological and
photometric properties.  The six panels show the redshift distribution
$z$, the absolute B magnitude $M_B$, the rest-frame color $(B-V)_0$,
the half-light radius $R_{hl}$, the bulge fraction $B/T$, and the
total asymmetry index, $R_T+R_A$\footnote{Definitions of DGSS
structural parameters may be found in Paper II of this series, Simard
\etal\ (2002).}.  Examination of Figure~\ref{hist} reveals that the 66
galaxies selected for chemical analysis are representative of the
entire DEEP sample in terms of their luminosities, redshift
distributions, sizes, and bulge fractions, but they are preferentially
bluer and more asymmetric than the sample as a whole.  This
disproportionate fraction of asymmetric galaxies with strong emission
lines may be understood either 1) as mergers which trigger star
formation and produce \HII\ regions, or 2) isolated galaxies dominated
by star-forming regions which give rise to an asymmetric morphology.
The possible systematic effects introduced by the selection criteria
will be discussed further below.

\section{Spectral Analysis}
\subsection{Emission Line Measurements}

Spectra taken on different observing runs with different slitmasks
enable us to combine multiple spectra for most DGSS galaxies. After
wavelength calibrating each spectrum, we combined the available
2-dimensional spectra to produce a master spectrum with higher
signal-to-noise.  We then extracted an optimal 1-D spectrum for
analysis.  The spectra are not flux calibrated.  We manually measured
equivalent widths of the emission lines present in each spectrum with
the IRAF SPLOT routine using Gaussian fits.  The [O~II] doublet, which
is visibly resolved, was fit with two Gaussian components and the sum
recorded.  Table~1 lists the equivalent widths and 
measurement uncertainties for
each line.  Where [O~III] $\lambda$4959 was below our nominal S/N
threshold of 8:1, we calculated the EW based on the strength of [O~III]
$\lambda$5007 assuming an intrinsic ratio of 3:1.  The reported
equivalent widths are corrected to the rest frame using

\begin{equation}
EW_{rest} = EW_{observed}/(1+z)
\end{equation}

\noindent Associated uncertainties are computed taking
into account both the uncertainty on the line
strength and the continuum level placement using

\begin{equation}
\sigma_{EW} = \sqrt{ {{1}\over{C^2}}\sigma_L^2 +
{{L^2}\over{C^4}}\sigma_C^2 },
\end{equation}

\noindent where $L$, $C$, $\sigma_L$, and $\sigma_C$ are the line and
continuum levels in photons and their associated $1~\sigma$ uncertainties.  We
determine $\sigma_C$ manually by fitting the baseline regions
surrounding each emission line multiple times.  We adopt
$\sigma_L=\sqrt{12}\times RMS$ where 12 is the number of pixels summed
in a given emission line for this resolution, and RMS is the root-mean-squared
variations in an adjacent offline region of the spectrum.  Using this
empirical approach, the stated uncertainties implicitly include errors
due to Poisson noise, sky background, sky subtraction, readnoise, and
flatfielding.  In nearly all cases, the continuum can be fit along a
substantial baseline region, so that $\sigma_C\ll\sigma_L$.

\subsection{AGN Contamination}

In the analysis that follows, accurate assessment of the chemical
abundances in the warm ionized medium requires that the observed
emission lines arise in H~II regions powered by photoionization from
massive stars.  Non-thermal sources such as active galactic nuclei
(AGN) often produce emission-line spectra that superficially resemble
those of star-forming regions.  AGN must be identified as such because
blindly applying emission-line metallicity diagnostics calibrated from
H~II region photoionization models will produce erroneous metallicities.

Traditionally, AGN can be distinguished from starbursts on the basis
of distinctive [N~II]/H$\alpha$, [S~II]/H$\alpha$, [O~I]/H$\alpha$, and
[O~II]/[O~III] line ratios (see Heckman 1980; Baldwin, Phillips, \&
Terlevich 1981; Veilleux \& Osterbrock 1987).  However,  some or most
of these diagnostic lines are unobservable in the current generation of
ground-based optical surveys at increasingly
higher redshifts.  Sometimes, as in this survey, only equivalent width
measurements are available.  With these limitations in mind,
Rola, Terlevich, \& Terlevich (1997) have explored using ratios of
$EW_{[O~II]}$ and $EW_{H\beta}$ as substitute diagnostics for identifying
AGN.

Figure~\ref{AGNtest} shows diagnostic diagrams,
$EW_{[O~II]}$/$EW_{H\beta}$ and $EW_{[O~III]}$/$EW_{H\beta}$ versus
$\log EW_{H\beta}$, for the identification of AGN and starforming
galaxies based on Rola \etal\ (1997).  The upper panels of
Figure~\ref{AGNtest} plot $EW_{[O~III]}$/$EW_{H\beta}$.  Filled symbols
denote the DGSS objects.  For a comparable sample of local star-forming
galaxies, we compiled three sets of spatially-integrated (i.e., global)
emission-line spectra from the literature.  The first set, consisting
of 22 objects, comes from the 55-object spectroscopic galaxy atlas of
Kennicutt (1992b).  For each galaxy with detectable (S/N$>$8:1)
[O~II]$\lambda$3727, [O~III]$\lambda$4959, [O~III]$\lambda$5007, and
H$\beta$ emission, we measured the emission-line fluxes and equivalent
widths.  Dereddened emission-line fluxes were computed by comparing the
observed $F_{H\alpha}/F_{H\beta}$ ratios to theoretical ratios
($I_{H\alpha}/I_{H\beta}$ = 2.75-2.85 for wide temperature range; here
we assume a fixed electron temperature of 12,000 K)\footnote{Direct 
measurements of the electron temperature
are not possible from these data since the temperature-sensitive 
[O~III] $\lambda$4363 line is not detected and the upper limits
do not provide useful constraints.}.  As a second local
sample, we selected the compilation of nearby field galaxies from
Jansen \etal\ (2000a,b; NFGS), adopting their published emission-line
fluxes and equivalent widths.   The Nearby Field Galaxy Survey is a
collection of 200 local ($z<0.04$) morphologically diverse galaxies
selected from the CfA I redshift catalog (Huchra 1983) which has the
virtue of being nearly complete to the photographic limits of the
survey.   See Jansen \etal\ (2000a) for a discussion of the differences
between the NFGS and K92 samples.  Briefly, the K92 objects have a
higher fraction of star-forming galaxies (objects with strong emission
lines) compared to the NFGS sample.  As a third local sample, we chose
the emission-line selected galaxies from the Kitt Peak National
Observatory Spectroscopic Survey (KISS; Salzer \etal\ 2000).  KISS is a
large-area objective prism survey of local ($z<0.09$) galaxies selected
by strong $H\alpha$ emission lines.

In Figure~\ref{AGNtest}, the distribution of DGSS objects is similar
to the distribution of KISS and K92 objects in having relatively large
$EW_{H\beta}$ and high $EW_{[O~III]}$/$EW_{H\beta}$ ratios.  In
comparison, the NFGS galaxies have lower mean $EW_{H\beta}$.  Open
symbols show a comparison sample of K92 local galaxies.  Skeletal
triangles denote the NFGS galaxies.  Dots indicate the KISS galaxies.
Filled symbols coded by redshift interval denote the new DGSS objects
presented here.  In the left panels we plot the raw equivalent width
data, while in the right panels we add 3 \AA\ to the $EW_{H\beta}$ as
a rough correction for underlying Balmer absorption in the stellar
population.  Solid lines indicate the regions described by Rola \etal\
as likely to contain AGN.  AGN typically occupy the upper left and
upper right sectors while starforming H~II galaxies typically fall in
the lower right and lower left regions. Many of the normal starforming
local galaxies from the K92 and NFGS compilations
also lie in parameter space ostensibly occupied by AGN, even in the
right panels where we have applied the 3\AA\ correction for
absorption.  It seems that, on the basis of equivalent width ratios,
there is considerable ambiguity regarding object classification.
We have chosen to be conservative and remove from our sample the 8
objects which fall in this region of the diagram given by
$EW_{[O~II]}$/$(EW_{H\beta})>5$ (dashed line) in the left (absorption
uncorrected) panels.  These 8 objects are colored magenta in
Figure~\ref{AGNtest}.  Even if some fraction of these 8 objects are
falsely flagged as AGN, their exclusion from the larger sample will
not have a significant impact on the conclusions drawn from the
56 remaining galaxies.

From examination of the spectra in Figure~1, we selected two additional
objects as probable AGN based on their anomalously large ratios of
[Ne~III] $\lambda3868$ to [O~II] $\lambda3727$.  Objects 092-7832 and
203-3109 do not stand out in Figure~\ref{AGNtest}, but have
$EW_{[Ne~III]\lambda3826}EW_{[O~II]\lambda3727}> 0.4$, compared to
$EW_{[Ne~III]\lambda3826}EW_{[O~II]\lambda3727}\sim0.1-0.2$ for normal
starforming objects in K92 Osterbrock (1989) notes the
presence of enhanced [Ne~III] as a common signature in AGN and LLINERs.  
Another
paper in this series will contain a more complete discussion of the AGN
population in the DGSS (Sarajedini \etal\ 2003; Paper XIII).

Nine of the lowest redshift objects have measurable $H\alpha$ and
[N~II] emission lines which can be used to place them on the more
traditional AGN diagnostic diagrams (e.g., Baldwin, Phillips, \&
Terlevich 1981; Veilleux \& Osterbrock 1987).  All of these objects
have [N~II]/H$\alpha$ and [O~III] line ratios consistent with normal
star-forming galaxies, [N~II]/H$\alpha\simeq0.1$.  These 9 objects
would have already been classified as star-forming galaxies on the
basis of our provisional [O~II]/$H\beta$ diagnostic in
Figure~\ref{AGNtest}.  Unfortunately, none of the possible AGN
identified above can be checked in this manner because the $H\alpha$
and [N~II] $\lambda6584$ lines are not included in the spectral
coverage.

In summary, 10 objects have been identified as possible AGN and
removed from the sample of 66 galaxies, leaving 56
intermediate-redshift field galaxies for further analysis.  Blindly
including the 10 probable AGN in the sample would reduce the mean
metallicity of DGSS galaxies (i.e., probable AGN preferentially
scatter to the metal-poor or overly-luminous side of the L-Z
relation).

Figure~\ref{select} shows in greater detail the distribution of
magnitude, color, restframe $EW_{H\beta}$ and $H\beta$ luminosity for
the 56 remaining galaxies compared to the original set of 276 DGSS objects
with emission lines in
the range $0.26<z<0.82$.  Symbols distinguish objects by redshift:
``low'' (14 objects; $0.26<z<0.40$), ``intermediate'' (19 objects;
$0.40<z<0.60$), and ``high'' (24 objects; $0.60<z<0.82$).  Filled
symbols denote galaxies selected for chemical analysis, while open
symbols denote the entire set of galaxies in each redshift interval.
Points with error bars denote the means and dispersions of each
sample.  The lower row shows $(B-V)_0$ color versus $M_B$.  Objects in
the lowest redshift bin exhibit a remarkable correlation between
luminosity and $(B-V)_0$ color, while the samples in the latter two
redshift bins do not.  The two highest redshift bins contain a higher
fraction of luminous blue objects which are absent in the low-redshift
bin.  In the high-redshift bin, the magnitude-limited nature of the
sample becomes obvious, as there are no galaxies fainter than
$M_B=-18$.  In all redshift bins, the selected galaxies are
preferentially those with the highest $EW_{H\beta}$ and the bluest
colors.  This effect is most significant in the highest redshift bin.
The correlation between blue magnitude and $H\beta$ luminosity in all
three redshift bins merely reflects the fact that the $H\beta$
luminosities were computed from $M_B$ and $EW_{H\beta}$.  No
extinction correction has been applied.  Put simply, more luminous
objects are capable of sustaining greater rates of star formation.
The distribution of equivalent widths indicates that the star
formation per unit luminosity is roughly similar for galaxies in the
two highest redshift intervals, but that the lowest redshift bin has
fewer objects with large EW.  This is most likely a volume effect
since there will be fewer extreme starburst objects in the smaller
volume at $0.26<z<0.40$.

\section{Analysis}
\subsection{Assessing the Physical and Chemical Properties of the Sample}

Oxygen is the most easily measured metal in \HII\ regions due to its
strong emission lines from multiple ionization species in the optical
portion of the spectrum.  Measurement of the oxygen abundance relative
to hydrogen is based upon the intensity ratio of the collisionally
excited [O~II]$\lambda$3727 and [O~III]$\lambda$4959,5007 lines
relative to Balmer series recombination lines (e.g., H$\beta$) using
the standard analysis techniques (see Osterbrock 1989).  Even when the
physical conditions of the ionized gas such as electron temperature and
density cannot be measured, the ratio of strong forbidden oxygen
emission lines can still provide a measure of the overall metallicity
of the gas (the so-called strong line $R_{23}$ ratio method; Pagel
\etal\ 1979; Kobulnicky, Kennicutt, \& Pizagno 1999, hereafter KKP).  
However, in the
DGSS and other large spectroscopic surveys, relative emission-line
intensities are often not measured.  Only equivalent widths are
available.  In a companion paper to this one, Kobulnicky \& Phillips
(2003) demonstrate that the ratio of emission-line equivalent widths,
$EWR_{23}$, is a quantity comparable to $R_{23}$ and is suitable for
measuring oxygen abundances.  We adopt the $R_{23}$ calibration of
McGaugh (1991) relating the ratio of [O~III] $\lambda\lambda$ 4959,
5007, [O~II] $\lambda$3727, and $H\beta$ to the oxygen abundance relative
to hydrogen, O/H.   Although multiple prescriptions have been proposed
in the literature, (reviewed in KKP and in
Figure~\ref{R23OH}, the exact choice is unimportant here since we are
only interested in {\it relative} abundances between local and distant
galaxy samples analyzed in the same manner.  Furthermore, the sample
includes only luminous galaxies ($M_B<-18$, with three exceptions) so
that, based on local analogs, they all are expected fall on the upper (metal-rich)
branch of the empirical strong-line calibrations.  The nine DGSS
galaxies with measurable [N~II] and H$\alpha$ lines all have
[N~II]/$H\alpha\gtrsim0.1$, confirming that these objects belong on the
metal-rich branch of the calibration.\footnote{Adopting the
less-plausible hypothesis that the DGSS galaxies lie on the metal-poor
branch of the double-valued strong-line $R_{23}$--$O/H$ relation would
require all the galaxies to have extremely low oxygen abundances,
$12+log(O/H)<8.0$.  Given that direct measurements of electron
temperatures and oxygen abundances in similarly luminous $z=0.4$
objects rule out the lower branch possibility (Kobulnicky \& Zaritsky
1999) we believe this upper-branch assumption to be generally valid.} 
To ensure that the NFGS and KISS comparison objects
are on the metal-rich branch, both local samples have been culled to
contain only objects with $F_{[N~II]\lambda6584}/F_{H\alpha}>0.15$.

In this paper, we compute oxygen abundances adopting the analytical
expressions of McGaugh (1991, 1998 as expressed in KKP) which are
based on fits to photoionization models for the metal-rich (upper)
branch of the $R_{23}$--O/H relation. In terms of the reddening
corrected line intensities, this relation is

\begin{eqnarray}
12+log(O/H) = 12 -2.939-0.2x-0.237x^2-0.305x^3-0.0283x^4-  \\
	y(0.0047-0.0221x-0.102x^2-0.0817x^3-0.00717x^4),
\end{eqnarray}

\noindent where

\begin{equation}
y \equiv\ \log(O_{32}) \equiv\ \log\Biggr(
	{{I_{[O~III]\lambda4959} + I_{[O~III] \lambda 5007}}
	\over{I_{[O~II]\lambda3727}}}\Biggr),
\end{equation}

\noindent and

\begin{equation}
x\equiv\ \log({R_{23}})\equiv\ \log\Biggr({{I_{[O~II] \lambda3727} + I_{[O~III]
\lambda4959} + I_{[O~III] \lambda 5007}}\over{I_{H\beta}}}\Biggr).
\end{equation}

\noindent Figure~\ref{R23OH} shows graphically the relation between
$R_{23}$, $O_{32}$ and oxygen abundance.  Alternative relations from
the literature are shown for comparison.  A star marks the Orion
nebula value (based on data of Walter, Dufour, \& Hester 1992) which
is in excellent agreement with the most recent solar oxygen abundance
measurements of $12+log(O/H)_\odot=8.7$ (Prieto, Lambert, \& Asplund
2001).

Kobulnicky \& Phillips (2003) show that strong line equivalent width ratios
are tightly correlated with flux ratios and can be used to obtain 
a metallicity-sensitive parameter akin to $R_{23}$.  
In our metallicity analysis we follow the 
prescriptions of Kobulnicky \& Phillips (2003).  
Using equivalenth widths, equations 5 and 6 become

\begin{equation}
y \equiv\ \log(EWO_{32}) \equiv\ \log\Biggr(
	{{EW_{[O~III]\lambda4959} + EW_{[O~III] \lambda 5007}}
	\over{EW_{[O~II]\lambda3727}}}\Biggr),
\end{equation}

\noindent and

\begin{equation}
x\equiv\ \log({EWR_{23}})\equiv\ \log\Biggr({{EW_{[O~II] \lambda3727} + EW_{[O~III]
\lambda4959} + EW_{[O~III] \lambda 5007}}\over{EW_{H\beta}}}\Biggr).
\end{equation}

\noindent As noted in Kobulnicky \& Phillips (2003),
the use of equivalent width ratios, rather than line flux
ratios, has the particular advantage of being less sensitive to the
(unknown) amount of extinction within the host galaxy, at least if the
reddening toward the gas and stars is similar.\footnote{Calzetti,
Kinney, \& Storchi-Bergmann (1994) present evidence that this assumption may
be invalid for some galaxies.}

Table~1 records the oxygen abundances, 12+log(O/H) derived from
tabulated emission-line equivalent widths using the $EWR_{23}$
prescription above.  The galaxies range between 12+log(O/H)=8.4 and
12+log(O/H)=9.0, with typical {\it random} measurement uncertainties of
0.03 to 0.10 dex.  An additional uncertainty of $\sim$0.15 dex in O/H,
representing uncertainties in the photoionization models and
ionization parameter corrections for empirical strong-line calibration,
should be added in quadrature to the tabulated measurement errors.
There may also be zero-point shifts required to make this calibration
directly comparable to standard solar oxygen abundances using
meteoritic or solar photosphere measurements.  The oxygen abundances
derived here should {\it not} be compared to the canonical solar
oxygen abundance (12+log(O/H)=8.89, Anders \& Grevesse 1989;
12+log(O/H)=8.69, Prieto, Lambert, \& Asplund 2001) which is measured
in a different manner.  Our oxygen abundances may, however, be
compared to local galaxies measured with the same empirical
calibration technique to establish relative metallicity trends.

Figure~\ref{zMBOH} shows the relationship between redshift, $M_B$, B-V
color\footnote{$B-V$ colors for the DGSS, KISS, and NFGS are taken directly
from published references.}, $EW_{H\beta}$, $L_{H\beta}$,
and oxygen abundance for the 56 objects in our sample.  Solid symbols
distinguish objects by redshift:  ``low'' (12 objects; $0.26<z<0.40$),
 ``intermediate'' (20 objects; $0.40<z<0.60$), and  ``high'' (24
objects; $0.60<z<0.82$).  Dots, crosses, and 3-pointed 
triangles designate the KISS, K92, and NFGS local galaxy
comparison samples.  Here, as in Figure~\ref{AGNtest},
the nearby comparison samples have been culled using the
same emission-line selection criteria as the DGSS galaxies.

The redshift--luminosity panels of Figure~\ref{zMBOH} show a
correlation.  As expected for a flux-limited survey, more high
luminosity objects and fewer low-luminosity objects are detected in
the highest redshift bin compared to the lowest redshift bin. The DGSS
galaxies are bluer than the mean NFGS galaxy, but consistent with the
bluest local galaxies.  The DGSS galaxies have, on average, larger
emission line equivalent widths.  The lower left panel compares the
oxygen abundance versus redshift.  The mean metallicity of the sample
increases with redshift across the three bins.  This trend is a
consequence of the combined effects of the flux-limited sample and the
luminosity-metallicity (L-Z) correlation for galaxies, pictured in the
far-right column.  The oxygen abundances of galaxies within each
redshift bin correlate strongly with blue luminosity and less strongly
with $H\beta$ luminosity.  There is no significant correlation between
color and metallicity or between equivalent width and metallicity.
The lower right panel reveals that the zero point of the L-Z
correlation is displaced toward higher luminosities for larger
redshifts.  DGSS galaxies lie predominantly on the upper envelope of
the local galaxy samples.  While most of the DGSS sample galaxies are
consistent with a luminosity-metallicity trend, the two least luminous
objects, 092-1375 and 172-1242 with $M_B\sim{-16.5}$, lie well off the
correlation if the empirical line-strength to-metallicity conversion
is blindly applied.  Based on studies of other galaxies of similar
luminosity ($M_B\simeq-16.5$; Skillman, Kennicutt, \& Hodge 1989), we
suspect that these two galaxies do not belong on the the upper branch
of the $R_{23}$--O/H relation.  They lie in the
``turn-around'' region between the upper and lower branches where the
strong-line calibration is particularly uncertain.  If these objects
lie on the lower branch of the $R_{23}$ calibration, then they have
metallicities near 12+log(O/H)=8.1.  Unfortunately neither of these
objects have $H\alpha$ or [N~II] detections, so there is no way to way
to discriminate between the two metallicity branches.  Due to their
metallicity ambiguity, we remove these two galaxies from any further
analysis.

\subsection{Comparison with Local Galaxies}

We turn now to a detailed comparison of the luminous and chemical
properties of the remaining DGSS sample to local galaxies and other
intermediate-redshift galaxies.  We select the same three sets of local
galaxies.  From the K92 sample, we use 22 comparison
objects.  From the photometry and spectroscopy presented in Jansen
\etal\ (2000a,b) we selected a sub-sample of 36 NFGS objects using the
same emission line criteria described above, including rejecting
objects fainter than $M_B=-17$.   From the KISS survey, we selected a
sub-sample of 80 objects using the same emission-line and luminosity
criteria which define our DGSS sample.
To ensure that the NFGS and KISS comparison objects
are on the metal-rich branch, both local samples have been culled to
contain only objects with $F_{[N~II]\lambda6584}/F_{H\alpha}>0.15$.

Figure~\ref{LZ} shows the luminosity-metallicity relation for DGSS
galaxies compared to the three local galaxy samples.  Filled symbols
denote DGSS galaxies in the redshift ranges $z=0.26-0.4$, $z=0.4-0.6$,
and $z=0.6-0.82$ as in Figure~\ref{zMBOH}.  Crosses, skeletal
triangles and dots denote the K92, the NFGS and the KISS local
samples, respectively.  The eight open squares show the best-measured
objects in the $0.6<z<1.0$ field galaxy study by Carollo \& Lilly
(2001; CL01).\footnote{Magnitudes have been converted from the
$H_0=50$, $\Omega_M=1.0$ ($q_0=0.5$) cosmology adopted in that study.
Oxygen abundances are computed using the published $R_{23}$ and
$O_{32}$ values from that work.} Open circles and triangles show the
$z=0.2-0.5$ objects from Kobulnicky \& Zaritsky (1999; KZ99) which
meet the same emission line selection criteria as the DGSS.  Open
stars are the high-redshift ($z>2$) galaxies from Kobulnicky \& Koo
(2000; KK00) and Pettini \etal\ (2001; Pe01).  Lines represent least
squares linear fits to the samples as shown in the key.  Fits to local
sub-samples are unweighted, whereas fits to the DGSS and high-$z$
sub-samples include the statistical uncertainties on O/H.  A
representative error bar in the lower right indicates the typical
statistical uncertainties of $\sigma(M_B)=0.2$ mag and
$\sigma(O/H)=0.12$ dex.  It is worth emphasizing here that the oxygen
abundances for each sample represented in Figure~\ref{LZ} have been
computed in an identical manner from measured emission line equivalent
widths.  No corrections for internal reddening have been made.  We
have adopted the recommendation of Kobulnicky \& Phillips (2003) and
corrected all local and DGSS galaxy $H\beta$ equivalent widths for
2 \AA\ of stellar Balmer absorption.
The absolute B magnitudes have been computed (or converted from
published values) assuming the $H_0=70$, $\Omega_M=0.3$,
$\Omega_\Lambda=0.7$ cosmology.  K-corrections have been applied to
the high-redshift samples based on multi-band rest-frame B-band and UV
photometry.  The various samples represented should be directly
comparable, although they might not be completely representative of
the entire population of galaxies at a given redshift.

Figure~\ref{LZ} reveals a striking variation in the zero point of the
best fit luminosity-metallicity correlation with redshift of the
sample.  The slope also varies with redshift.  Although considerable
scatter exists within a given sub-sample, {\it the more distant
galaxies are markedly brighter at a fixed oxygen abundance.}
Equivalently, galaxies of a given luminosity are systematically more
metal-poor at higher redshifts.  Dashed lines show the linear least
squares fits to the local NFGS samples using both O/H and $M_B$ as the
dependent variable.  The heavy dashed line shows the linear bisector
of the two fits and has the form: $12+log(O/H)=-0.086~M_B + 7.14$.
The best fit to the local KISS galaxies in the upper panel is similar
with $12+log(O/H)=-0.12~M_B + 6.59$.  For the $z\sim0.7$ DGSS
galaxies, we find $12+log(O/H)=-0.16~M_B + 5.19$.

Figure~\ref{LZ} underscores and extends the trend from
Figure~\ref{zMBOH} by including galaxies at higher and lower
redshifts.  The most dramatic differences between the local and
higher-redshift DGSS samples occur at the lower luminosities.  At
luminosities fainter than $M_B=-19$, the mean galaxy in the
$z=0.60-0.82$ bin is a factor of 2 (0.2--0.3 dex) more metal poor than
the NFGS or KISS local samples. At high luminosities, there is
considerable overlap between all subsamples up to $z=0.8$.  At
$M_B=-21.5$, the mean galaxy in the $z=0.60-0.82$ subset is perhaps
only 0.1 dex more metal poor than the NFGS or KISS local samples,
although the paucity of local galaxies prohibits a detailed
comparison.  The lack of DGSS galaxies in the lower right corner of
the plot (low luminosity but high metallicity) represents a real
offset from the local relation, and is not due any selection effect we
can identify.  Galaxies that might occupy this locus near $M_B=-19$,
$12+log(O/H)=8.7$ would have measurable emission lines and line
ratios, $R_{23}$, identical to the more luminous galaxies with
$M_B=-21.5$.  Although the ratios of [O~III] and [O~II] line strengths
to H$\beta$ line strengths drop with increasing metallicity, the
sample selection process would not preferentially exclude objects with
low [O~III]/H$\beta$ (high metallicity) in favor of those with high
[O~III]/H$\beta$ (low metallicity).  If such low-luminosity,
high-metallicity objects existed, they would have been detected and
included in the analysis.  In our selection of objects from the DGSS
database, we found only 3 objects with measurable $H\beta$ but
immeasurably weak [O~III].  In all those cases, poor sky subtraction
is clearly the cause of the weak [O~III].  Furthermore, there is no
correlation between $EW_{H\beta}$ and O/H (see Figure~\ref{zMBOH}), as
might be expected if selection effects had preferentially removed
objects with weak line strengths from the sample or if the
corrections for Balmer absorption dominated the errors in a systematic
way.

We investigated other sources of systematic or selection effects by
examining the residuals in the L-Z relation as a function of other
fundamental galaxy parameters.  Using unweighted linear least-squares
fits to the KISS, NFGS, and DGSS samples separately, we computed
residuals for each sample and searched for correlations with galaxy
color, $H\beta$ luminosity, size, and equivalent width.  Only
$EW_{H\beta}$ is significantly correlated with L-Z residuals, and only
for the DGSS galaxies.  Figure~\ref{LZresidew} plots galaxy
$EW_{H\beta}$ versus luminosity residuals, $\delta{M_B}$, from a
best-fit linear relation in the L-Z plane.  There is no correlation
for the KISS sample, or for the NFGS sample.  There is a significant
correlation for the intermediate- and high-redshift DGSS galaxies
which is driven by the handful of galaxies with $EW_{H\beta}>20$ \AA.
This may plausibly be understood as a sign that the galaxies with
large $EW_{H\beta}$ are undergoing the strongest episodes of star
formation, and are correspondingly offset to the bright side of the
L-Z relation.

Comparing local galaxies having moderate $EW_{H\beta}$ to the rare
DGSS galaxies with uncommonly large $EW_{H\beta}$ values could lead to
misleading conclusions about the nature of distant versus local populations.
With this in mind, we have constructed a new version of
Figure~\ref{LZ} using only galaxies with $EW_{H\beta}<20$ \AA\ where
the DGSS and local galaxies should be comparable.  
Figure~\ref{LZlim} shows only those DGSS and local KISS, NFGS, and K92
galaxies with $EW_{H\beta}<20$\AA.  Although the number of DGSS data
points are fewer, best fit lines to each sample show that the basic
conclusion remains the same.  DGSS galaxies in the high-redshift bin
are as luminous as local galaxies at the luminous end of the L-Z
relation, while the least luminous DGSS galaxies are, on average,
$\sim$1---2 mag more luminous than local galaxies of comparable
metallicity.  Again, this result does not appear to be due to some
kind of selection effect since the DGSS sample analyzed here spans the
entire range of luminosities from the entire DGSS sample in each
redshift interval, i.e., we are not simply selecting the brightest
objects from each redshift interval, as shown by Figure~\ref{select}.

Figure~\ref{select} hints at another kind of selection effect which
may lead to erroneous conclusions regarding differences between DGSS 
and local galaxy samples.
All of the DGSS hgalaxies have blue colors with $B-V<0.6$ while
only a small fraction of the local galaxies are as blue.  The DGSS
samples are likely to contain a larger fraction of extreme star-forming
galaxies with younger light-weighted stellar populations.
As a second means of examining selection/population differences between the
local and DGSS samples, we have constructed an alternative version of
Figure~\ref{LZ} using only local galaxies with the bluest colors.
Figure~\ref{LZlimBV} shows only those DGSS and local KISS and NFGS
galaxies with $B-V<0.6$.  Although the number of NFGS and KISS data
points are fewer, best fit lines to each sample show that the basic
conclusion remains the same.  DGSS galaxies in the high-redshift bin
are as luminous as local galaxies at the luminous end of the L-Z
relation, while the least luminous DGSS galaxies are, on average,
$\sim$1---2 mag more luminous than local galaxies of comparable
metallicity. 

As an even more stringent comparison, we considered only DGSS, NFGS,
and KISS galaxies with $EW_{H\beta}<20$ \AA\ {\it and} $B-V<0.6$. 
We further culled the local samples to remove objects with 
$EW_{H\beta}<10$. The
basic conclusion remains the same even for this very restrictive
comparison designed to mitigate the color and EW differences between
the DGSS and local samples.  DGSS galaxies in the high-redshift bin
are as luminous as local galaxies at the luminous end of the L-Z
relation, while the least luminous DGSS galaxies are, on average,
$\sim$1---2 mag more luminous than local galaxies of comparable
metallicity.

A straightforward interpretation of Figure~\ref{LZ} and
Figure~\ref{LZlim} is that the least luminous emission line galaxies
{\it have undergone considerable luminosity evolution, $\Delta
M_B\simeq1-2$ mag, from $z\sim0.7$ (7 Gyr ago) to the present and as
much as $\Delta M_B\simeq3-4$ mag since $z\sim3$ (11 Gyr ago).}  This
amount of brightening is consistent with the results of Ziegler
\etal\ (2002) who found an evolution of 1-2 mag in the Tully-Fisher
relation for the least massive galaxies in their $z=0.1-1$ sample.
An alternative way of understanding
our result is to say that field galaxies
with $M_B\sim-18$ have experienced a measurable degree of metal
enrichment (a factor of 2) at constant luminosity since $z\sim0.8$.
The arrows in Figure~\ref{LZlim} indicate the evolution in the L-Z
plane caused by constant star formation, passive evolution, metal-poor
gas inflow, and star formation bursts and/or galaxy mergers.  Some
combination of these processes are responsible for evolving the
$z=0.6-0.8$ galaxies into the region occupied by today's $z=0$
galaxies.

The apparent evolution in the luminosity-metallicity relation seen in
Figures~\ref{LZ} and \ref{LZlim} stands in contrast to the conclusions
of Kobulnicky \& Zaritsky (1999) and Carollo \& Lilly (2001) who found
little or no evolution in the redshift ranges $0.1<z<0.5$ and
$0.5<z<1.0$, respectively.  Open squares in Figure~\ref{LZ} show that
the CL01 objects fall at the high-luminosity end of the
luminosity-metallicity correlation where the local and ``distant''
samples overlap.  A comparison of the CL01 objects to the local NFGS
sample would find that they are likely to be drawn from the same
distribution, and hence the conclusion of CL01 is appropriate given
the restricted nature of that sample.  The new conclusion based on
Figure~\ref{LZ} is made possible by the addition of larger numbers of
galaxies and lower-luminosity galaxies which greatly extend the
luminosity and metallicity baselines.  Although the Kobulnicky \&
Zaritsky (1999) sample included a broader range of luminosities, the
least luminous objects were at the lowest redshifts where offsets from
local galaxies are least pronounced.  Furthermore, Kobulnicky \&
Zaritsky (1999) used a local comparison sample that included only a
restricted subset of of galaxies from the Kennicutt (1992a,b) spectral
atlas.  Objects from this sample show increased scatter toward
low-metallicities at fixed luminosity due, mostly, to the errors
introduced by variable, unknown amounts 
of stellar Balmer absorption.  The increased scatter in the
local sample masked the trend which has now become apparent in
Figure~\ref{LZ}.  This evolution of the L-Z relation between $z=0$ and
$z=0.7$ supports the conclusions of Kobulnicky \& Koo (2000), Pettini
\etal\ (2001) and Mehlert \etal\ (2002) who found high redshift
galaxies to be overly luminous for their metallicity at $2.2<z<3.4$.

\section{Discussion}
\subsection{Modeling the Evolution in the L-Z Plane }

The addition of chemical information on galaxies at earlier epochs
provides a new type of constraint on theories of galaxy formation and
evolution.  If local effects such as the gravitational potential and
``feedback'' from supernova-driven winds are the dominant regulatory
mechanisms for star formation and chemical enrichment, then the L-Z
relation might be nearly independent of cosmic epoch.  The
semi-analytic models of Kauffman (1996) and Somerville \& Primack (1999)
show little or no evolution in the L-Z relation with epoch, although
model-dependent prescriptions for stellar feedback and galactic winds
are poorly constrained and may have large impacts on the chemical and
luminous evolution.  Based on the current understanding of cosmic
evolution, that the volume-averaged star formation rate was higher in
the past (Pei \& Fall 1995; Madau 1996), and that the overall
metallicity in the universe at earlier times was correspondingly lower,
we might expect galaxies to be considerably brighter at a given
metallicity (i.e., luminosity evolution) if there was more
primordial gas available in the universe to fuel star formation.  A high- or
intermediate-redshift galaxy sample ought to be systematically
displaced from the local sample in the luminosity-metallicity (L-Z)
plane {\it if individual galaxies reflect these cosmic evolution
processes,} i.e., if they have an exponentially declining star
formation rate since some epoch of formation, as commonly assumed, and/or
if lower-mass galaxies evolve at different rates than high-mass
galaxies.  Comparison of the data from Figure~\ref{LZ} to some
simple galaxy evolution models can help distinguish between these
possibilities.

In a simple evolutionary model, galaxies begin as parcels of
gas which form stars and produce metals as the gas fraction decreases
from 100\% to 0\%.  The B-band luminosity\footnote{Here, we consider
only the B-band luminosity-metallicity relation.  Since young stars
dominate the B-band luminosity, we expect that much of the scatter of the 
B-band L-Z relation is due to extinction and fluctuations in the star
formation rate.  Ideally, a rest-frame R-band or I-band
luminosity-metallicity relation would be a more sensitive tool for
evolution studies, since it will be less sensitive to these
effects and eliminate the need for the color correction discussed in
the previous section.} is proportional to the star formation rate, which
may vary as a function of time, but is generally believed to be
proportional to the mass or surface density of the remaining gas
(Kennicutt 1998).  For a galaxy that evolves as a ``closed box'',
converting gas to stars with a fixed initial mass function and chemical
yield, the metallicity is determined by a single parameter: the gas
mass fraction, $\mu=M_{gas}/(M_{gas}+M_{stars})$.   The metallicity,
$Z$, is the ratio of mass in elements heavier than He to the total
mass, and is given by

\begin{equation}
Z= Y ln(1/\mu) \label{mu},
\end{equation}

\noindent where $Y$ is the ``yield'' as a mass fraction.  A typical
total metal yield for a Salpeter IMF integrated over 0.2--100
$M_\odot$ is $Y=0.012$ by mass (i.e., 2/3 the solar metallicity of 0.018;
see Pagel 1997, Chapter 8).\footnote{We have assumed the solar oxygen
abundance to be $12+log(O/H)_\odot\simeq$8.7 based on the new solar
oxygen abundance determination of Prieto, Lambert, \& Asplund (2001).
This is 0.1-0.2 dex lower than the oft-adopted Anders \& Grevesse
(1989) value, but resolves the discrepancy between solar and Orion
nebula oxygen abundances.  For the Orion nebula, $R_{23}=0.70\pm0.03$,
$O_{32}\simeq0.2\pm0.1$ (Walter, Dufour, \& Hester 1992), which, given
the calibration cited (Equation 3), leads to $12+log(O/H)=8.7\pm0.02$,
in excellent agreement with the Prieto, Lambert, \& Asplund (2001)
measurement.} A total oxygen yield for the same IMF would be $Y_{O}=0.006$.
Effective yields in many local galaxies seem to range from solar to
factors of several lower, (Kennicutt \& Skillman 2001; Garnett 2002),
implying either that the nucleosynthesis prescriptions (e.g., Woosley
\& Weaver 1995) are not sufficiently precise, 
that assumptions about the form of the
initial mass function are incorrect, or that metal loss is a
significant factor in the evolution of galaxies. Garnett (2002) finds
that oxygen yields of  $Y_{O}=0.001$ to  $Y_{O}=0.014$ 
among local irregular and spiral
galaxies are correlated with galaxy mass, suggesting an increasing
amount of metal loss among less massive galaxies.

High-velocity winds capable of producing mass loss are observed in
local starburst galaxies (e.g., Heckman \etal\ 2000) and in
high-redshift Lyman break galaxies (Pettini \etal\ 2001), but the
actual amount of mass ejected from galaxies is difficult to estimate.
Simulations indicate that winds may be metal-enriched, such that metals
are lost from galaxies more easily than gas of ambient composition
(Vader 1987; De Young \& Gallagher 1990; MacLow \& Ferrara 1999).
Recent Chandra observations of a local dwarf starburst galaxy,
NGC~1569, provide the first direct evidence for metal-enhanced
outflows (Martin, Kobulnicky, \& Heckman 2002). Thus, the closed-box
models are probably not appropriate and more realistic models
including selective metal loss are required.

In order to compare the luminous and chemical evolution of galaxies in
the L-Z plane to theoretical expectations, we ran a series of
\textsc{P\'egase2} (Fioc \& Rocca-$\!$Volmerange 1999) models. 
\textsc{P\'egase2} is a galaxy evolution code
which allows the user to specify a range of input
parameters including the initial mass function (which we take to be
the Salpeter value, -2.35, between 0.1 and 120 \mo, following Baldry
\etal\ 2002), and the chemical yields (we assign Woosley \& Weaver
1995 B-series models for massive stars). The effective
metal yield of these models is $Y=0.016$.   For other critical
parameters such as the star formation rate and the timescale on which
the galaxy is assembled, we explored a range of formulations. We
included an inclination averaged extinction prescription as
implemented in \textsc{P\'egase2}, but extinction changes the model B
magnitudes negligibly by only 0.2 mag.  Our goal in each case was to
reproduce the qualitative behavior of Figure~\ref{LZ} where
high-luminosity galaxies at $z=0.7$ occupy a similar locus as
high-luminosity galaxies at $z=0$ and where low-luminosity $z=0.7$
galaxies are displaced toward brighter magnitudes and lower
metallicities relative to their $z=0$ counterparts. 

We tried a range of SFR prescriptions, including constant and
exponentially declining star formation rates, for preconstituted
galaxies of a fixed gas reservoir.  In each case, both the initial star
formation rate and total gas reservoir available had to be carefully
tuned in an arbitrary manner to produce gas exhaustion at just the
right epoch to account for the similarity of the $z=0.7$ and the 
local relations at the highest luminosities.  By the time the evolution
in the L-Z plane slowed when the gas supply neared exhaustion at
$\mu<0.05$, the
model galaxies all had several times the solar metallicity, in conflict
with the $z=0$ and $z=0.7$ observations.

We next tried a more physically motivated prescription, where the star
formation rate is proportional to the mass of accumulated gas (or the
gas density; Schmidt 1959; Kennicutt 1998).  Galaxies are built
by infall of primordial-composition gas with an infall rate
that declines with an exponential timescale, $\tau_{infall}$,
as implemented in \textsc{P\'egase2},

\begin{equation}
M(t) = M_{tot} { {e^{-t/\tau_{infall} } \over{\tau_{infall}} }}.
\end{equation}

\noindent 
We chose two representative gas infall timescales of 1 Gyr and 5 Gyr
and two representative total gas masses of $10^{10}$ \mo\ and
$10^{11}$ \mo.   A 
significant constraint on the models is the position of $z\simeq3$ Lyman break
galaxies which lie 2-4 magnitudes brighter (0.5 dex more metal-poor)
than the local L-Z relation.  If the present-day emission-line samples
are at all representative of the evolved descendants of the $z=0.7$
and $z\simeq3$ galaxies,\footnote{Most of the DGSS objects studied
here have optical morphologies (see Figure~\ref{SpecIm1}) that
qualitatively resemble disks, and their bulge-to-disk ratios are $B/T
<< 1$.  Even if the current levels of star formation decline, their
evolutionary descendants are likely to be systems with substantial
disk components.  Thus, it is plausible to think that the descendants of the
$z\sim0.7$ DGSS galaxies are disk-like star-forming galaxies today,
similar to the KISS and NFGS objects plotted in Figure~\ref{LZ}.} then
a successful model must reproduce the rapid rise in metallicity and
luminosity in the first 2-3 Gyr (as evidenced by the Lyman break
galaxies), and it must reproduce the lack of significant L-Z evolution
during the past 5-7 Gyr as evidenced by the similarity of the most
luminous $z=0.7$ galaxies to the present-day L-Z relation.
Furthermore, a successful model should indicate that the least luminous
(least massive?) galaxies exhibit more pronounced luminosity and/or
metallicity evolution during the last 7 Gyr than the more luminous
objects in the sample.  This means that the rate of evolution of the
L-Z relation must be rapid during the first few Gyr of a galaxy's
formation, and then drop dramatically as a galaxy approaches the
present-day L-Z relation.  A slowing of luminous and/or chemical
evolution of a galaxy must physically correspond either 1) to
cessation of star formation and ultimately to depletion of the gas
available to fuel star formation, or 2) to continued inflow of
metal-poor gas which fuels star formation and dilutes the composition
of the ISM so that a galaxy remains fixed in L-Z space.

Figure~\ref{LZevp2} shows evolutionary tracks of galaxies in the L-Z
plane based on the infall \textsc{P\'egase2} models.  
The metallicity by mass, $Z$, is marked
along the upper x-axis along with the corresponding gas mass fraction
given by the models (equivalent to the gas mass fractions
predicted by Equation~\ref{mu} using a yield of 0.016) as a comparison.
Pentagons in Figure~\ref{LZevp2} mark the model galaxies at ages of 1,
2, 4, 8, and 12 Gyr (from left to right).  Dashed lines denote
models with $\tau_{infall}=1$ Gyr and solid lines denote models with
$\tau_{infall}=5$ Gyr.  The upper set of curves shows galaxies with
$M_{tot}=10^{11}$ \mo\ while the lower curves show galaxies with
$M_{tot}=10^{10}$ \mo.  These models produce the rapid rise
in metallicity in the first few Gyr.
These model masses are in good agreement with
observational estimates which place lower limits on the masses of
$z\simeq3$ Lyman break galaxies at $few\times10^{10}$ \mo\ (Kobulnicky
\& Koo 2000; Pettini \etal\ 2001) and indicate star formation rates of
$\sim$50 \mo/$yr^{-1}$. 
However, these models
over-enrich the ISM at late times after 8 Gyr compared to the data.
They predict gas mass fractions that are as low as 1\% and metallicities
as high as 3 times solar (Z=0.054) after 12 Gyr, at odds with
the observed properties of nearby galaxy samples.  Lowering the model 
star formation rates would delay the chemical enrichment until later times
and lower the luminosity at any given time.  

As an experiment to produce models that better resemble the data, we
introduced a metal yield efficiency parameter $\eta_Z$ which reduces
the effective yield of the models.  Physically, $\eta_Z$ may represent
the fraction of metals retained in galaxies during episodes of
wind-driven metal loss, or it may represent a correction to the
theoretical yields from stellar populations.  Figure~\ref{LZevp}
shows the same models and data as Figure~\ref{LZevp2}, except that the
metal yield has been reduced by a factor of 2
($\eta_Z=0.5$).\footnote{This factor is not incorporated into the
\textsc{P\'egase2} models in any self-consistent manner.  Here we have
simply reduced the final metallicity of the gas at each timestep by
the factor $\eta_Z=0.5$. }  The $M_{tot}=10^{11}$ \mo\ model with
$\tau_{infall}=1$ Gyr is now in better agreement with the $z=0.7$ and
$z=0$ data at ages of 8-12 Gyr. However, the same model for
$M_{tot}=10^{10}$
\mo\ becomes too metal-rich at ages of 8-12 Gyr compared to the $z=0$
data.  The $\tau_{infall}=5$ Gyr provides a better fit to the $z=0.7$
and $z=0$ low-luminosity galaxies at these ages.  Models with even
longer infall timescales would lie even further to the left of the solid
pentagons.

The evolution of the slope and zero point of the L-Z
relation in Figure~\ref{LZevp} is consistent with the idea that
low-mass field galaxies assemble on longer timescales and require
longer to reach high metallicities compared to their massive
counterparts in the field.\footnote{Some galaxies, particularly
dwarfs, may never exhaust their gas by the normal star formation
process and may never reach high metallicities if the gas reservoir is
cut off or removed prematurely by ram pressure stripping in a cluster
environment or by galactic winds.}  Alternatively, the same results may
be achieved if
the low-mass galaxies begin their assembly process at a
later cosmic epoch.  There are not enough constraints in the models
or the data to distinguish between these possibilities.
In the context of the cold dark
matter paradigm, the objects with the largest initial overdensities
form first (e.g., Blumenthal \etal\ 1984; Navarro, Frenk, \& White,
1995).  Whether these first objects grow to become massive galaxies
depends on the environment in which they are conceived.  Initial
galaxy cores which collapse in the vicinity of larger concentrations
of baryonic matter grow to become the massive galaxies at the centers
of today's galaxy clusters.  Objects that form in regions of lower
cosmological overdensity would accrete gas on longer timescales and,
if isolated from external influences, would proceed more slowly
through the star formation and chemical evolution process.  A longer
formation timescale for lower mass field galaxies observed in the DGSS
at $z\sim0.7$ can plausibly explain the observed evolution in the L-Z
relation. Such a scenario, whereby galaxies assembly timscales
vary with environment, is similar to that
proposed by Sandage, Freeman, \& Stokes (1970) to explain the differences
along the Hubble sequence.  

An additional free parameter not easily incorporated into the
\textsc{P\'egase2} models is the role of starbursts on timescales of few Myr 
that occur in a galaxies evolving on long (Gyr) timescales according
to the model prescriptions above. Studies of local galaxies show that
their star formation histories are not smooth but instead are punctuated by
periods of enhanced activity.  Such short-term enhancements in the
star formation rate of galaxies otherwise evolving smoothly can
elevate the luminosity and emission line equivalent widths by factors
of several.  The observed emission line equivalent widths in DGSS
galaxies average 10-20 \AA, but some exceed 50!  These
characteristics would be better fit by superimposing a burst of star
formation of $\sim10^{6-7}$ \mo\ on the model galaxies.  The impact of
instantaneous bursts on the evolution in the L-Z plane would be to
create spikes along the model tracks toward higher luminosities.  The
amplitude of the spikes would scale with the amplitude of the burst,
and could conceivably be several magnitudes and would persist for few
100 Myr. Given the morphological 
evidence for ongoing mergers among the DGSS sample,
evidenced by inspection of Figure~1, it seems certain that short-term
enhancements in the star formation rate are responsible for the more
extreme luminosities and emission line equivalent widths among the
sample.  Examination of the \textsc{P\'egase2} models used here shows
that as the $EW_{H\beta}$ declines from 50 \AA\ to 10 \AA, the B-band
luminosity drops by 0.7 mag. If we were to try to correct the
magnitudes of the DGSS samples by some factor to account for their
larger equivalent widths compared to the local samples, this factor
would be less than 0.7 mag, not enough to account for the larger 1-2
mag offsets observed in the L-Z relations.

\subsection{A Hypothetical L-Z Relation with Redshift}

Figure~\ref{LZevt} shows schematically what the evolution in the
L-Z plane may look like as a function of redshift.  
Lines at $z=0$ and $z=0.7$ are drawn to match the formal fits to the 
data in these redshift bins.  The other hypothetical lines
are drawn to indicate a possible evolutionary trend in the L-Z relation
which is consistent with the (currently minimal) observational
constraints.  This schematic shows more clearly than Figure~\ref{LZevp}
the more pronounced evolution for low-mass galaxies compared to high
mass galaxies and illustrates the increasingly rapid evolution
at redshifts beyond $z=2$.  Figure~\ref{LZevt} may be
approximated by an analytical function which expresses
metallicity in terms of luminosity, $M_B$, and redshift, $z$ as

\begin{equation}
12+log(O/H)=\left[ (-0.095 z )-0.13\right] M_B + (-2.31 z + 6.18).
\end{equation}
  
\noindent In terms of the metallicity by mass, $Z$, an equivalent
expression is

\begin{equation}
log Z=\left[ (-0.21 z )-0.10\right] M_B + (-4.9 z -6.9),
\end{equation}

\noindent where $\log Z_\odot=-1.74$.

\noindent Because this schematic is based upon observations 
of luminous ($M_B<-18$), metal-rich ($ 12+log(O/H)>8.3$)
galaxies, it may be inapplicable to the least luminous, metal-poor
galaxies where the behavior of the L-Z relation may change (e.g., 
Melbourne \& Salzer 2002).  Extrapolation beyond the 
current luminosity and redshift data is imprudent.

\section{Conclusions}

Observations of star-forming galaxies from the Deep Extragalactic
Evolutionary Probe (DEEP) survey of Groth Strip galaxies in the
redshift range $0.26<Z<0.82$ show a correlation between B-band
luminosity and oxygen abundance, like galaxies in the local universe.
The L-Z relations for $z<$0.8 intersects with the L-Z relation at
$z=0$ for the most luminous ($M_B<-20.5$) DGSS and local
galaxies. There appears to be very little evolution in the L-Z plane
among these most luminous galaxies over the last 8 Gyr.  However, less
luminous (and by implication, possibly less massive) galaxies 
are 1-2 mag more luminous at $0.6<z<0.8$ than $z=0$ galaxies of
similar metallicity.  Said another way, galaxies of comparable
luminosity are 0.1-0.2 dex more metal-rich at $z=0$ compared to
$z=0.7$. The least luminous galaxies appear to undergo considerable
evolution in the L-Z plane.  This differential evolution can be
explained in one of two ways.  Either less luminous (less massive?) 
galaxies are assembled on longer timescales than the super $L^*$
galaxies, or they begin their formation process at a later cosmic
epoch.  Evolution of galaxies in the L-Z plane over the redshift range
$z=3$ to $z=0$ appears to be a combination of fading and chemical
enrichment for plausible models.  Loss of up to 50\% of the oxygen
appears to be necessary to avoid over-enriching the galaxies at the
observed redshifts.  Because the local comparison sample has been been
selected on the same emission-line basis as the intermediate-redshift
samples, accounting for differences in color, this result is
consistent with evolution in the luminous and chemical properties of
star-forming galaxies over the last 4-8 Gyr.

How does this result compare to conclusions about the evolution of the
luminosity function (LF)?  Lin \etal\ (1999) use early CNOC2 data on
$>2000$ galaxies over the redshift range $0.12<z<0.55$ to conclude
that, among late-type galaxies (the ones most comparable to the DGSS
objects selected here), changes in the LF are most consistent with
{\it density} evolution, and little or no {\it luminosity evolution}.
However, Cohen (2002), based on studies in the Hubble Deep Field North
vicinity over the redshift range $0.01<z<1.5$, concluded that
emission-line objects show {\it moderate luminosity evolution} and
{\it little density evolution}.  Interpreting Figure~5 of Lin \etal\
(1999) as luminosity evolution suggests that, between $z\simeq0.70$
and $z\simeq0.20$, late-type galaxies with $M_B=-19$ fade by $\simeq$1
mag, while galaxies with $M_B=-21$ fade by $\simeq0.2$ mag.  This is
qualitatively and quantitatively consistent with the amount of fading
required for the $z=0.7$ DGSS galaxies to match the local L-Z relation
in the models in Figure~\ref{LZevp}.

These results imply that $z\sim0.6-0.8$ DGSS galaxies must have
significantly higher (20\%-80\%) gas mass fractions than comparably
luminous local galaxies.  Such a difference in the mean gas mass
fractions of galaxies would, in principle, be observed with instruments
such as the proposed radio-wave square kilometer array.  New
optical and near-IR spectroscopic
observations can test the schematic predictions of Figure~\ref{LZevt}
by measuring oxygen abundances of objects fainter than $M_B\sim-20.5$
in the critical redshift range $z=1$ to $z=2$ (i.e., by filling in the
low-luminosity side of Figure~\ref{LZevp}).

\acknowledgments

We thank John Salzer for insightful conversations and for providing
the KISS data in electronic form, Shiela Kannappan for a helpful
discussion about the NFGS, Matt Bershady and James Larkin for
scientific inspiration.  Detailed reading by an anonymous referee
greatly improved this manuscript.  H.~A.~K was supported by NASA
through grant
\#HF-01090.01-97A awarded by the Space Telescope Science Institute
which is operated by the Association of Universities for Research in
Astronomy, Inc. for NASA under contract NAS 5-26555 and by NASA
through NRA-00-01-LTSA-052.  This work was also made possible by NSF
grants AST95-29098 and AST00-71198 and NASA/HST grants AR-07532.01,
AR-06402.01, and AR-05801.01.

\clearpage

\begin{figure}
\figcaption[SpecIm1.ps] {Unfluxed spectra for 66 emission-line galaxies culled
from the DEEP survey of the Groth Strip using the criteria discussed in
Section~2 along with their HST F814W images.  Object are ordered by
redshift, as in Table~1, with the 10 possible or probable AGN
at the end.  Markings identify major emission lines of the [O~II]$\lambda$3726/29
doublet, [Ne III] $\lambda3868$, H$\beta$, [O~III]$\lambda\lambda$4959,5007, 
and, where applicable, $H\alpha$.  Other
positive features are generally residuals from night sky lines which
could not be completely subtracted.  Each panel shows
two different spectra, obtained with ``red'' and ``blue'' grating settings.
Seven-digit strings in the upper left 
give the DGSS identification number.  Rectangles on 
the images show spectroscopic slit positions.   (All 10 panels appear 
in the electronic edition only).
\label{SpecIm1} }
\end{figure}

\plotone{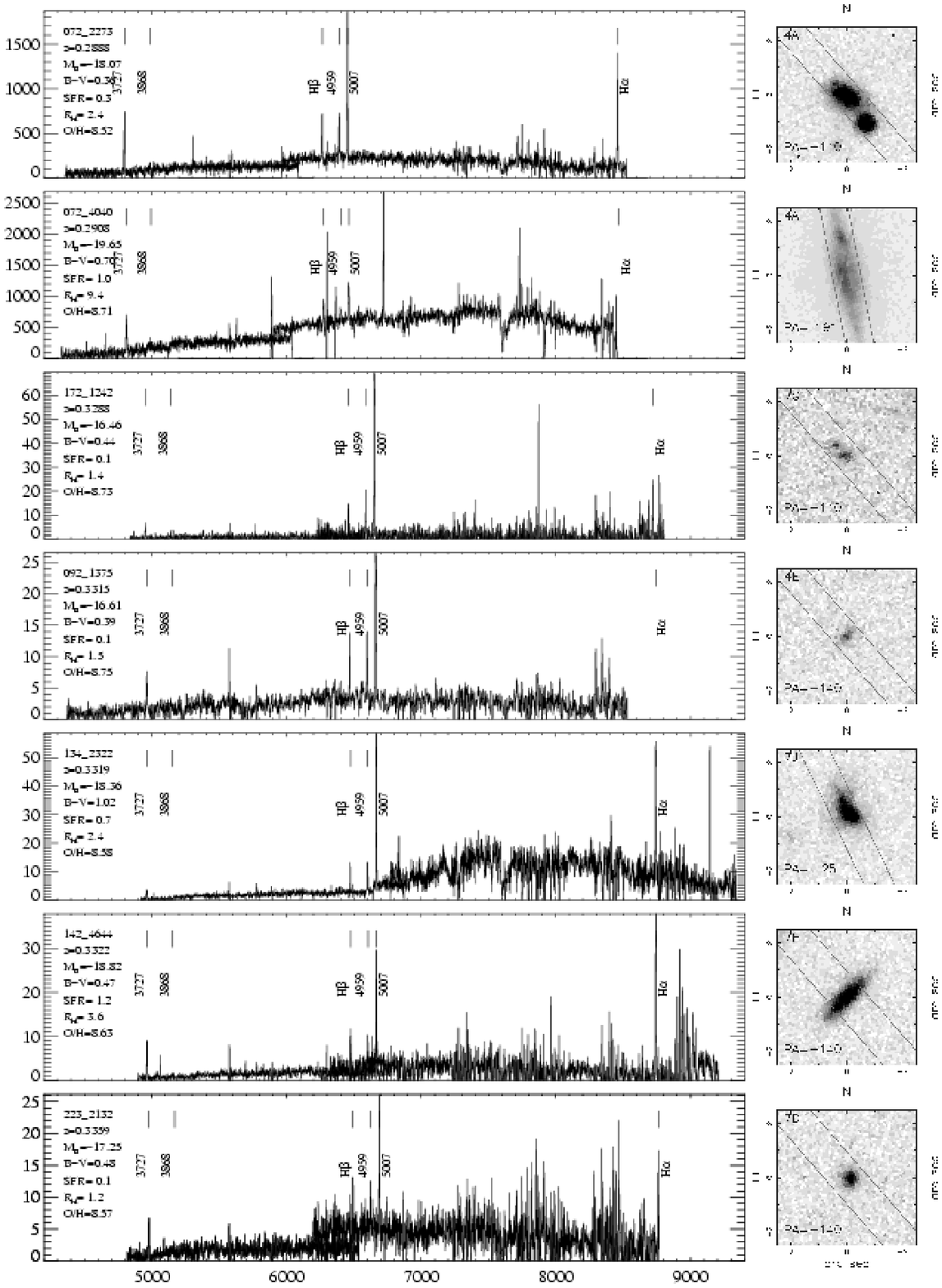}
\plotone{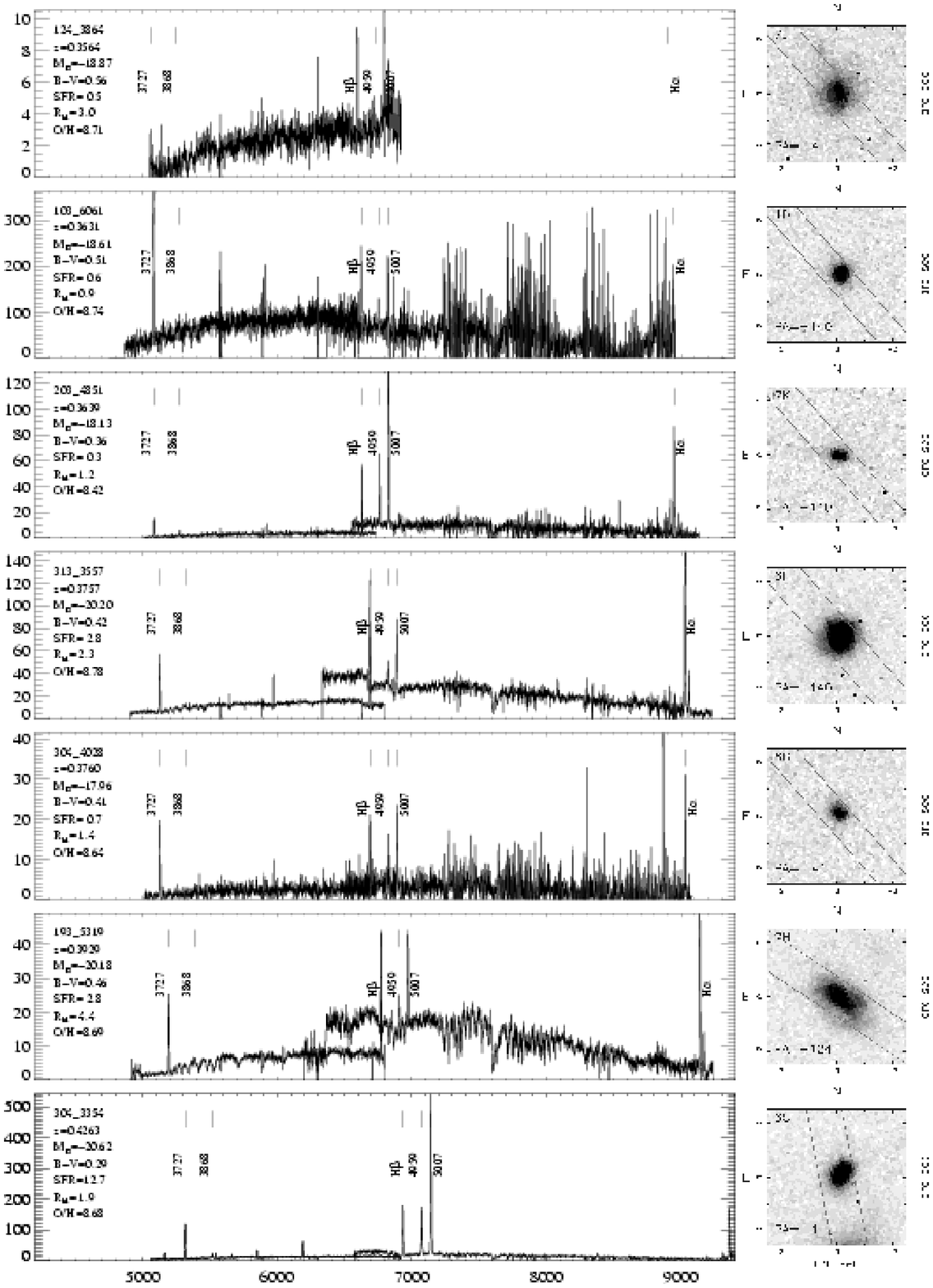}

\clearpage

\begin{figure}
\plotone{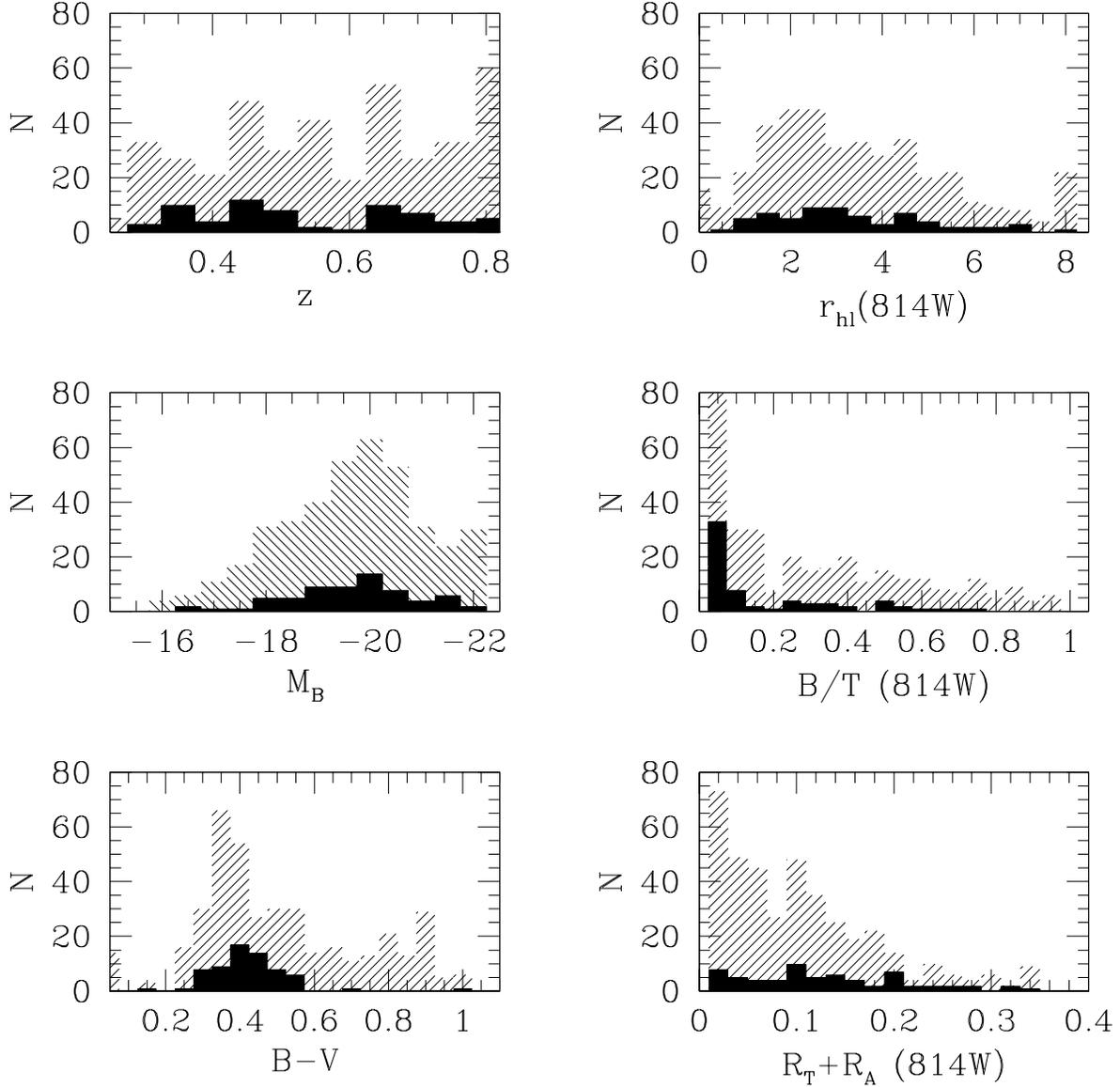}
\figcaption[hist1] {Histogram of 66 galaxies selected for chemical
analysis (filled) compared to the total 398 objects in the survey with
$0.26<z<0.82$.  We show the distribution as a function of redshift,
$M_B$, B-V color, half-light radius $R_{hl}$, bulge fraction
$F_{bulge}$, and asymmetry index, $R_T+R_A$. 
This figure demonstrates that
galaxies selected as suitable for chemical analysis
are representative of the larger DGSS sample in
the same redshift range
in terms of their redshift distributions,
sizes, luminosities and bulge fractions.  
However, the 66 selected galaxies preferentially 
have the bluest B-V colors and the most asymmetric 
morphologies.  \label{hist} }
\end{figure}

\clearpage

\begin{figure} 
\plotone{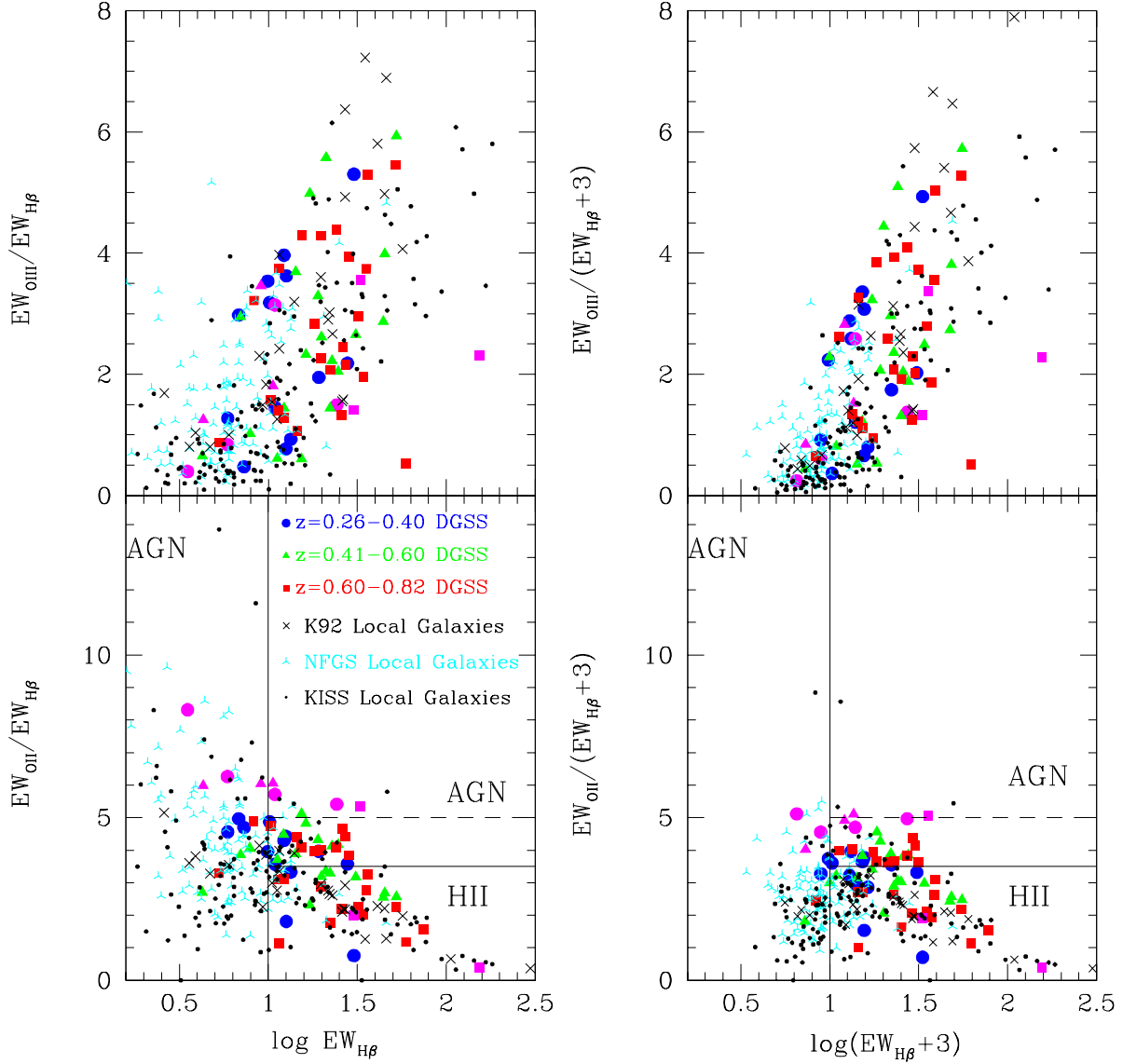}
\figcaption[AGNtest.ps] {Diagnostic diagrams, $EW_{[O~II]}/EW_{H\beta}$
versus $\log EW_{H\beta}$, for the identification of AGN and starforming
galaxies based on Rola \etal\ (1997).  
The full NFGS and KISS local samples are shown.
AGN typically occupy the upper
left and upper right sectors while starforming H~II galaxies typically
fall in the lower right regions, as divided by solid
lines.  The lower left sector is occupied by
both normal galaxies with low star formation rates and some AGN.  Open
symbols show a comparison sample of local galaxies drawn from Kennicutt
(1992b) and the NFGS (Jansen \etal\ 2000).  Filled symbols coded by
redshift denote the new objects presented here.  On the left we plot
the raw equivalent width data, while on the right we add 3 \AA\ to the
$EW_{H\beta}$ as a rough correction for underlying Balmer absorption in
stars.   On the basis of this diagram we identify and
remove from the sample 8 DGSS galaxies (magenta-colored) with
$EW_{[O~II]}/EW_{H\beta}>5$ (dashed lines) as possible AGN or objects with
large, unconstrained amounts of stellar Balmer absorption.  
\label{AGNtest} } \end{figure}

\clearpage

\begin{figure} 
\plotone{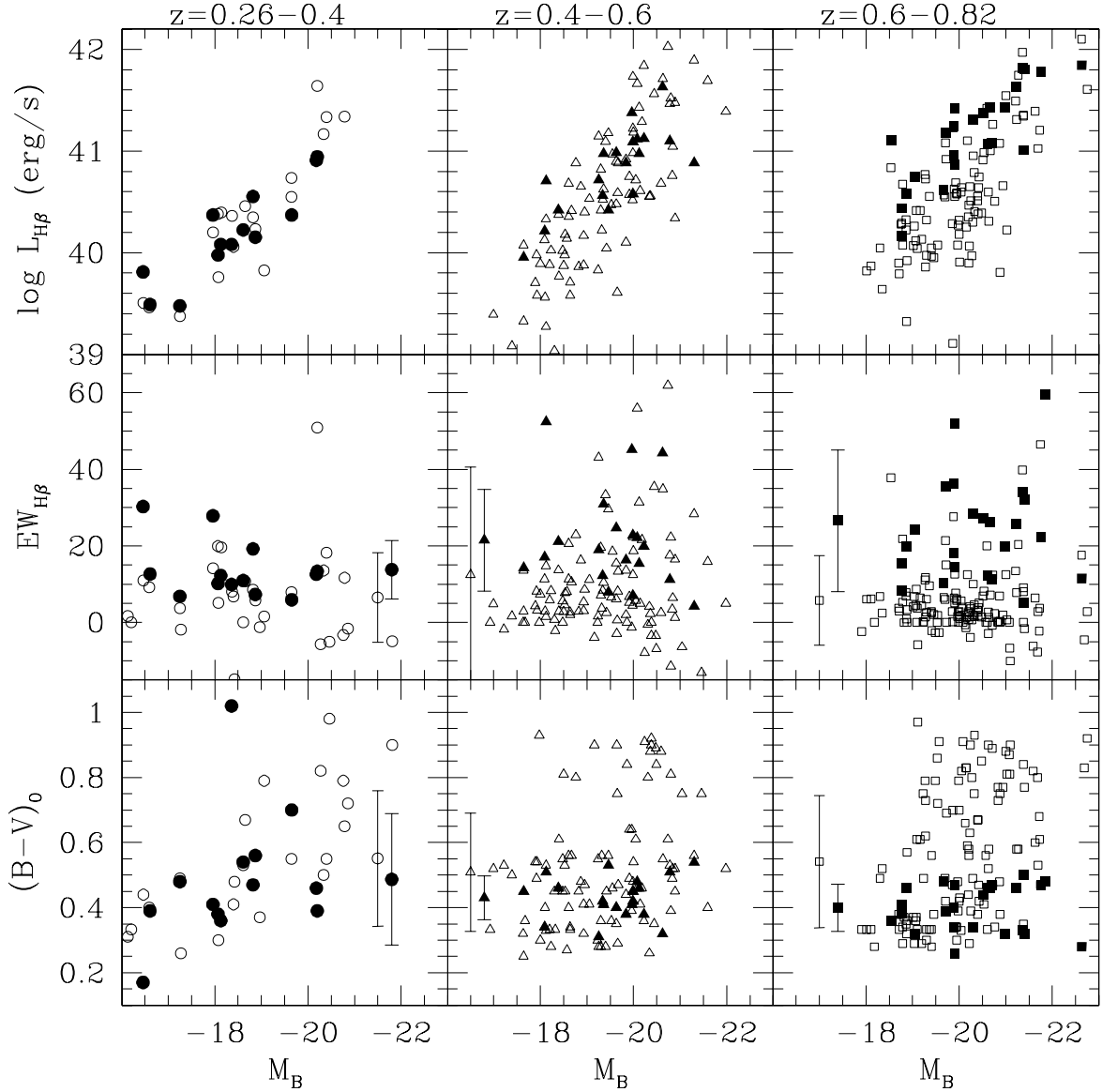}
\figcaption[select.ps] {A comparison of the luminosity, color, and
emission line strength for 56 galaxies selected for chemical analysis
(solid symbols) to the 276 DGSS galaxies with
emission lines (open symbols) selected from among
the 398 DGSS galaxies in the $0.26<z<0.82$ range.  Error bars shows means and dispersions
for the subsamples.  Objects are binned by redshift.  Galaxies
selected for analysis in the lowest redshift bin are representative of
the DGSS objects in that redshift interval.  However, galaxies selected
for analysis from the highest redshift bins are biased toward
the bluest colors and highest emission line equivalent widths.
The correlation between $M_B$ and $L_{H\beta}$ luminosity is an artifact
produced by computing $L_{H\beta}$ from $M_B$ and $EW_{H\beta}$.
\label{select} } \end{figure}

\clearpage

\begin{figure}
\plotone{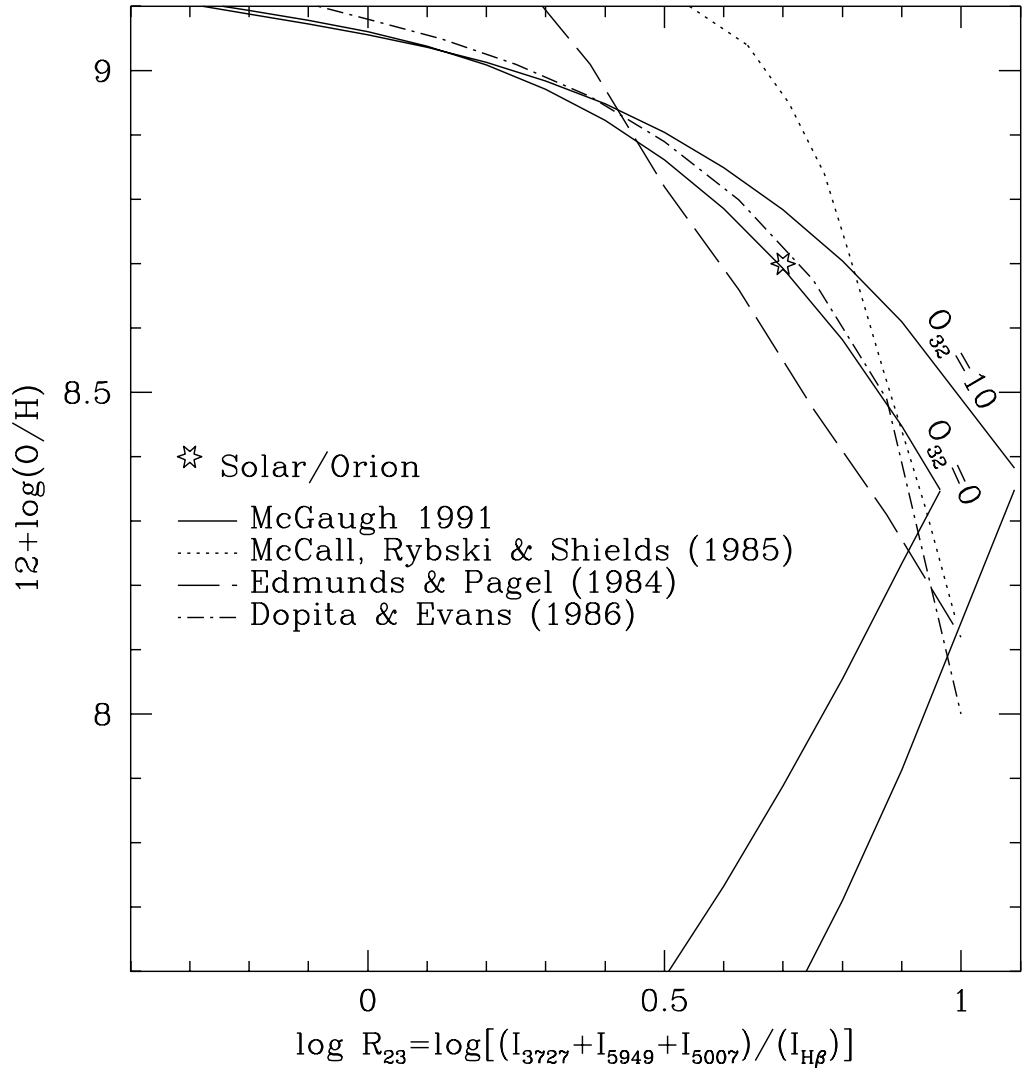}
\figcaption[R23OH.ps] {
Relation between the emission line ratio, $\log R_{23}\equiv
\log [ (I_{3727}+I_{4959}+I_{5007}) / (I_{H\beta}) ]$,
and oxygen abundance, 12+log(O/H) for
a variety of calibrations from the literature.
We adopt an analytical expression from
McGaugh (1991) which takes into account the ionization parameter
as measured by the $O_{32}$ ratio.  
A star marks the Orion nebulae oxygen abundance  
(Walter, Dufour, \& Hester 1992) which is consistent with
the solar abundance, based on the 
recent measurement by Prieto, Lambert, \& Asplund (2001).  
\label{R23OH} }
\end{figure}

\clearpage

\begin{figure}
\plotone{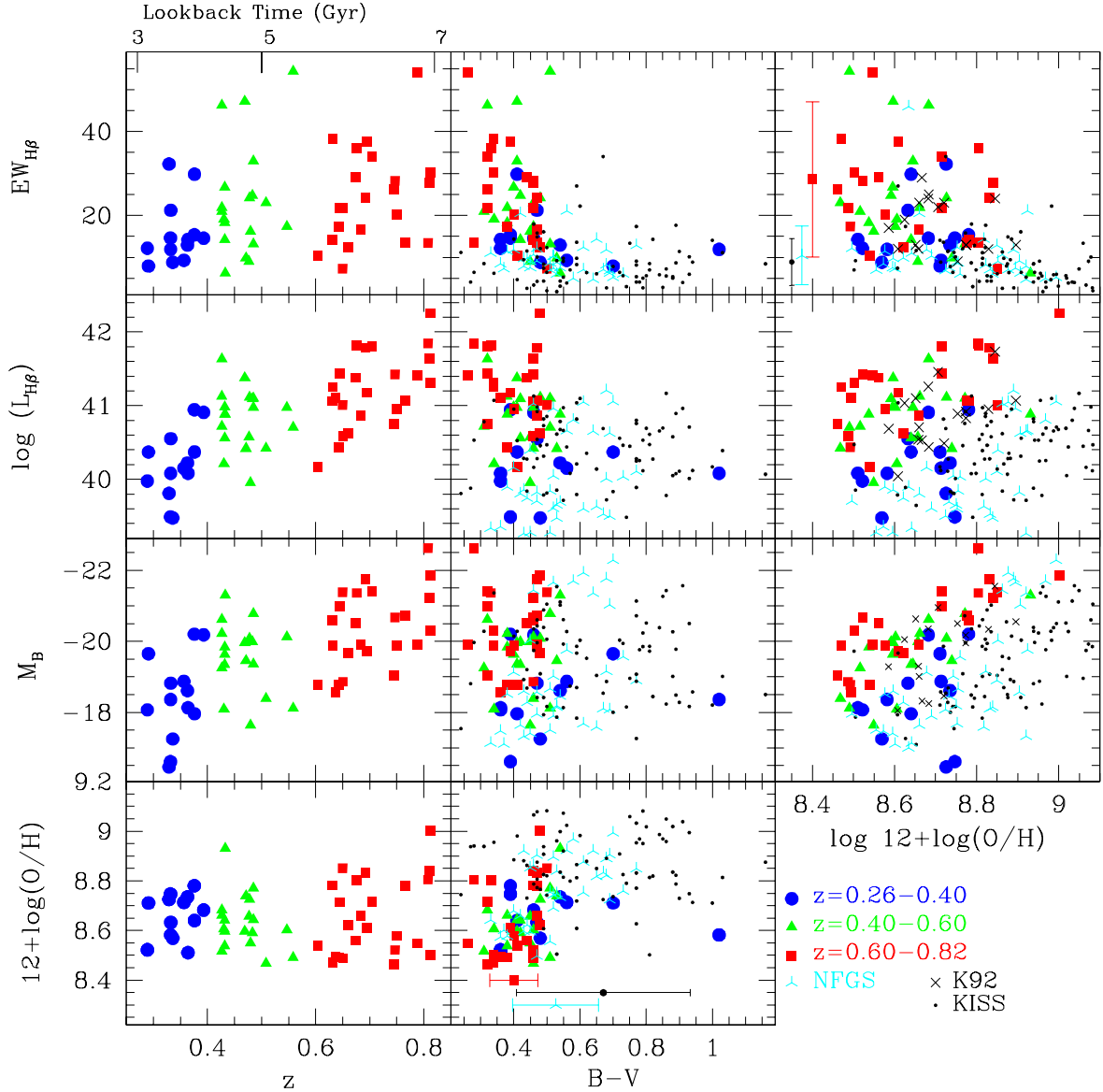}
\figcaption[zMBOH.ps]{Relation between redshift, $M_B$, 
B-V color, $EW_{H\beta}$, $L_{H\beta}$, and 12+log(O/H)
for 56 DGSS galaxies with subsets of the NFGS, K92, and KISS
samples chosen to match the DGSS emission line ratio criteria.  
Points with error bars in some panels compare the means and dispersions
in color and $EW_{H\beta}$ for the DGSS, NFGS, and KISS samples.
DGSS galaxies in the highest redshift bin are preferentially
bluer with higher star formation rates and larger emission line equivalent
widths.  Each redshift interval exhibits
a correlation between metallicity and both blue and $H\beta$
luminosities, but with different zero points.  There is no 
significant correlation
between metallicity and color, or between metallicity and emission line
equivalent width. \label{zMBOH} }
\end{figure}

\clearpage

\begin{figure}
\plotone{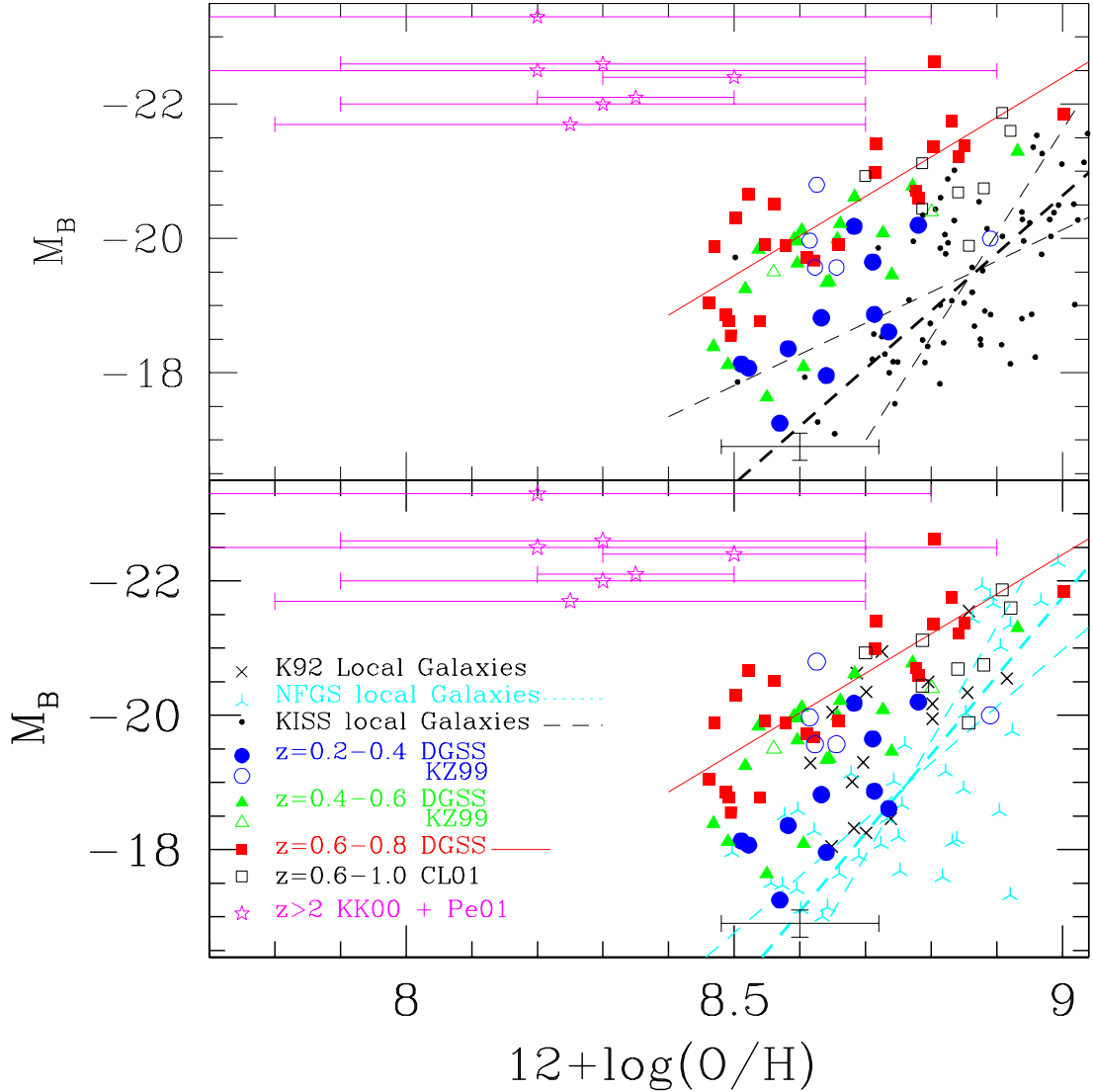}
\figcaption[LZ.ps] {Luminosity-metallicity (L-Z) relation including
samples of local $z<0.1$ galaxies from the NFGS, KISS, and Kennicut
(1992b) catalogs, with intermediate-redshift DGSS objects from this
study (filled symbols) and the Carollo \& Lilly (2001) and Kobulnicky
\& Zaritsky (1999) compilations (open symbols).   A representative
error bar is shown.  Stars denote high-redshift objects (Kobulnicky \&
Koo 2000; Pettini \etal\ 2001) with error bars representing the
full permitted range of metallicities ($\sim3\sigma$).  The
intermediate-redshift L-Z relation is consistent with the local L-Z
relation for luminous galaxies, but diverges toward lower metallicity.
Residuals from the local L-Z relation are most pronounced for
low-luminosity DGSS galaxies in the highest redshift bin.  This is
consistent with the hypothesis that the most massive objects complete
their evolution more rapidly, and the least massive objects evolve more
slowly.  \label{LZ} } \end{figure}

\clearpage

\begin{figure} 
\plotone{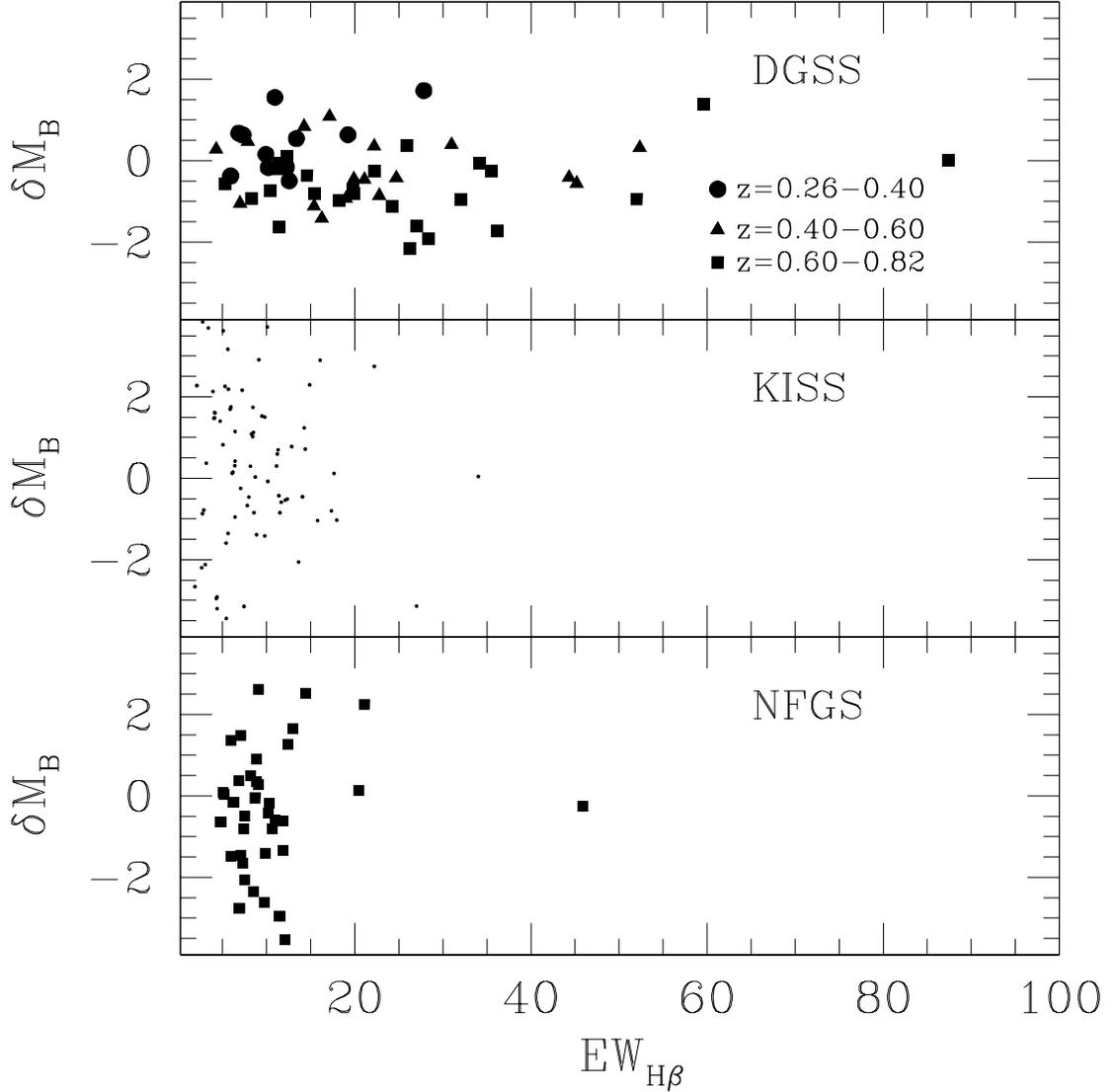}
\figcaption[LZresidEWHB.ps] {Galaxy $EW_{H\beta}$ versus luminosity residuals,
$\delta{M_B}$, from a best-fit linear relation in the L-Z plane for the
DGSS, KISS, and NFGS samples.  There is no correlation between $EW_{H\beta}$
and L-Z residuals for the local samples.   There is a weak
correlation among the DGSS galaxies, driven mostly by the few galaxies
with extremely large  $EW_{H\beta}$. We searched for
other parameters, including galaxy color and size, which might correlate with
L-Z residuals and explain some of the dispersion in the
L-Z relation, but no significant correlations were found.
\label{LZresidew} } \end{figure}

\clearpage

\begin{figure}
\plotone{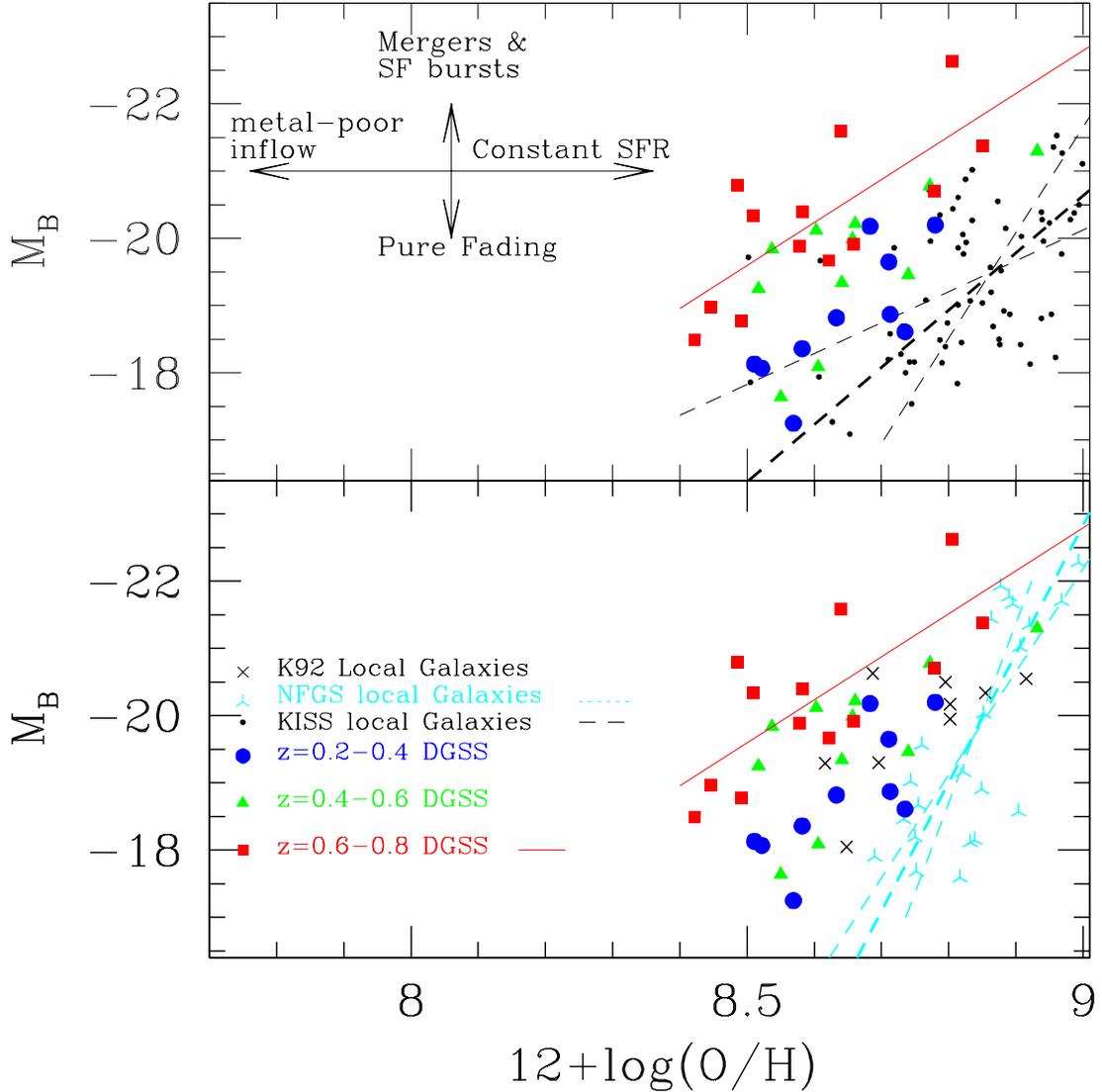}
\figcaption[LZ.ps] {Luminosity-metallicity (L-Z) relation 
showing only local and DGSS galaxies with
 $EW_{H\beta}<20$ \AA.  Here again, 
the most luminous DGSS galaxies are as luminous as the local
galaxies of comparable metallicity, while
the least luminous DGSS galaxies are $\sim1-2$ mag
brighter than local galaxies of comparable metallicity.
The schematic at left indicates the evolution in the L-Z plane
caused by constant star formation, passive evolution, 
metal-poor gas inflow, and star formation bursts and/or 
galaxy mergers.  Some combination of these processes are
responsible for evolving the $z=0.6-0.8$ 
galaxies into the region occupied by today's
$z=0$ galaxies.  \label{LZlim} } \end{figure}

\clearpage

\begin{figure}
\plotone{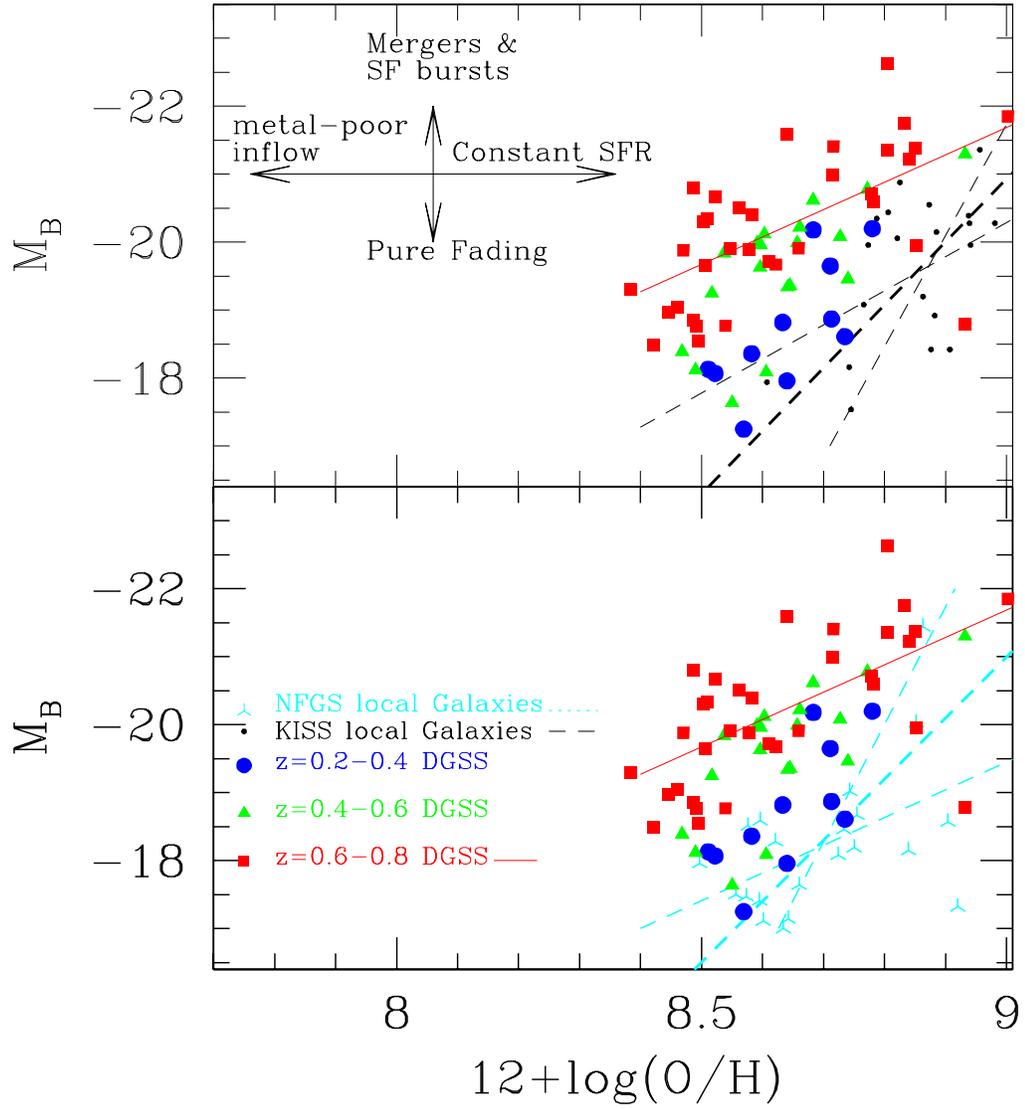}
\figcaption[LZlimBV.ps] {Luminosity-metallicity (L-Z) relation 
showing only local and DGSS galaxies with
 $B-V<0.6$ \AA.  Here again, 
the most luminous DGSS galaxies are as luminous as the local
galaxies of comparable metallicity, while
the least luminous DGSS galaxies are $\sim1-2$ mag
brighter than local galaxies of comparable metallicity.
 \label{LZlimBR} } \end{figure}

\clearpage

\begin{figure} 
\plotone{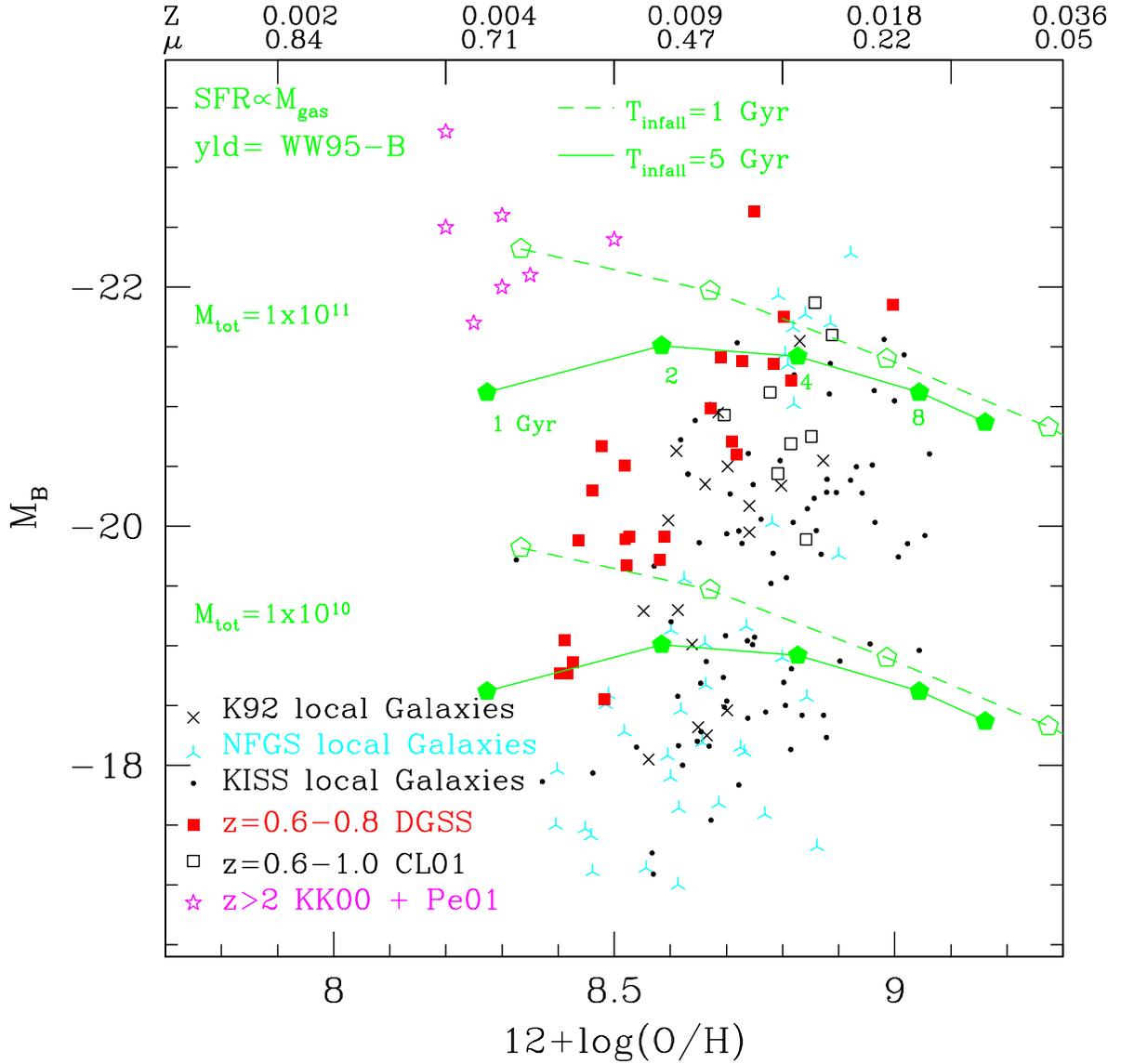}
\figcaption[LZevp2.cps] 
{L-Z relation with a subset of the symbols from
Figure~\ref{LZ}.  Lines and pentagons show the evolution of a model galaxy
with a star formation rate proportional to the gas mass where the
galaxy is built by exponentially-decreasing infall of primordial gas
with infall timescale of 1 Gyr and 5 Gyr.  Pentagons denote galaxies at
2, 4, 8, and 12 Gyr.  Galaxy masses of $10^{10}$ \mo\
and $10^{11}$ \mo\ are shown.  The top of the figure shows the
corresponding gas metallicity, $Z$, and mass fraction, $\mu$, assuming 
a closed-box scenario for an effective yield of 0.012 from Weaver \& 
Woosley (1995) series B models. 
The relative offset of low-mass $z=0.6-0.8$ galaxies from
the local L-Z relation can be understood as the result of a longer gas
infall timescale or a later formation epoch
for low-mass galaxies compared to high-mass galaxies.
\label{LZevp2} } \end{figure}
\clearpage

\begin{figure} 
\plotone{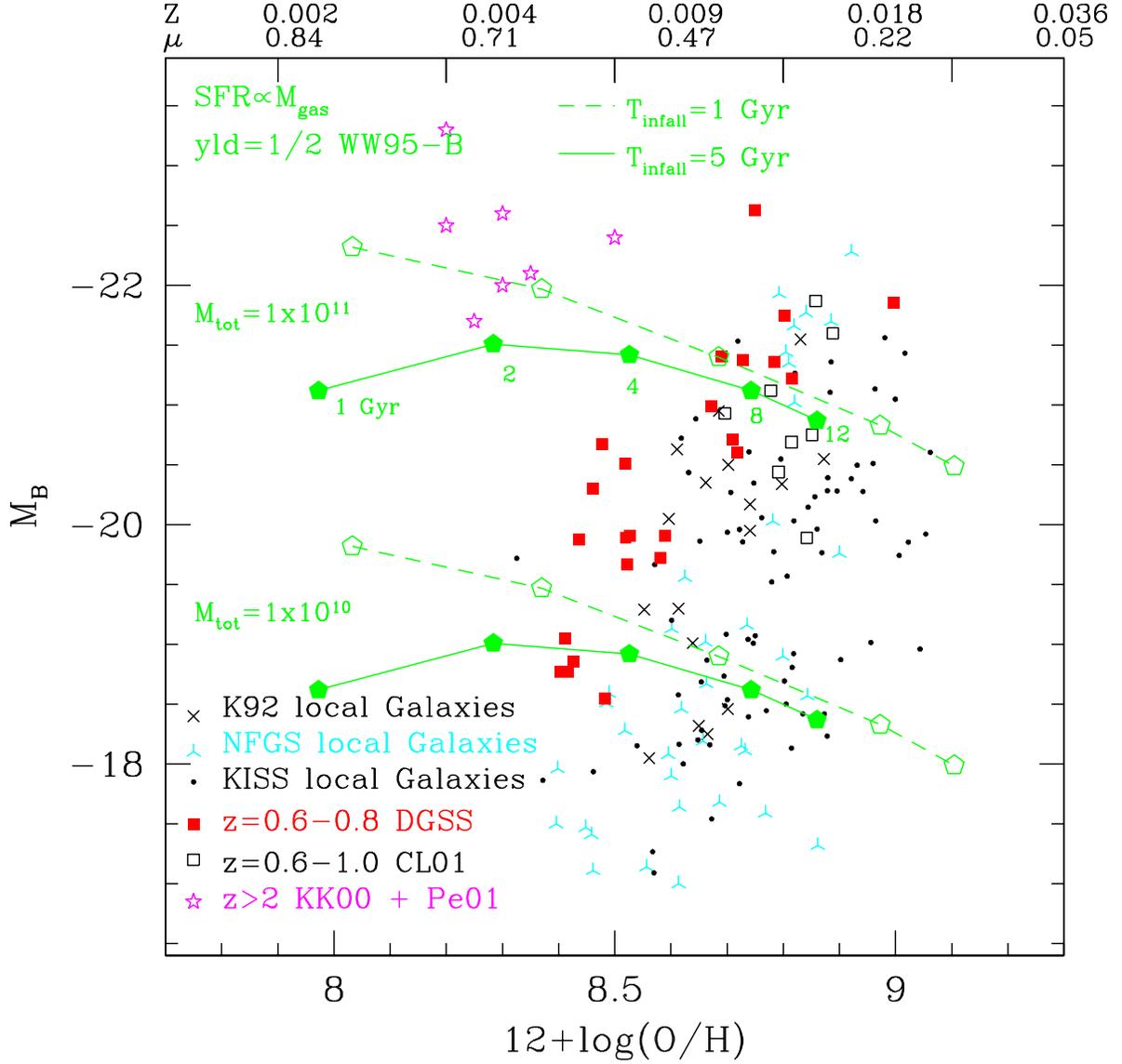}
\figcaption[LZevp.cps] 
{Comparison of data to evolutionary models in the L-Z plane similar to
Figure~\ref{LZevp2}, except that the effective yield of the models
has been arbitrarily reduced by a factor of 2.  
The metallicities of the reduced-yield models do not
rise as quickly, do not overproduce metals at
late times, and are in better agreement with the data.  
\label{LZevp} } \end{figure}

\clearpage

\begin{figure} 
\plotone{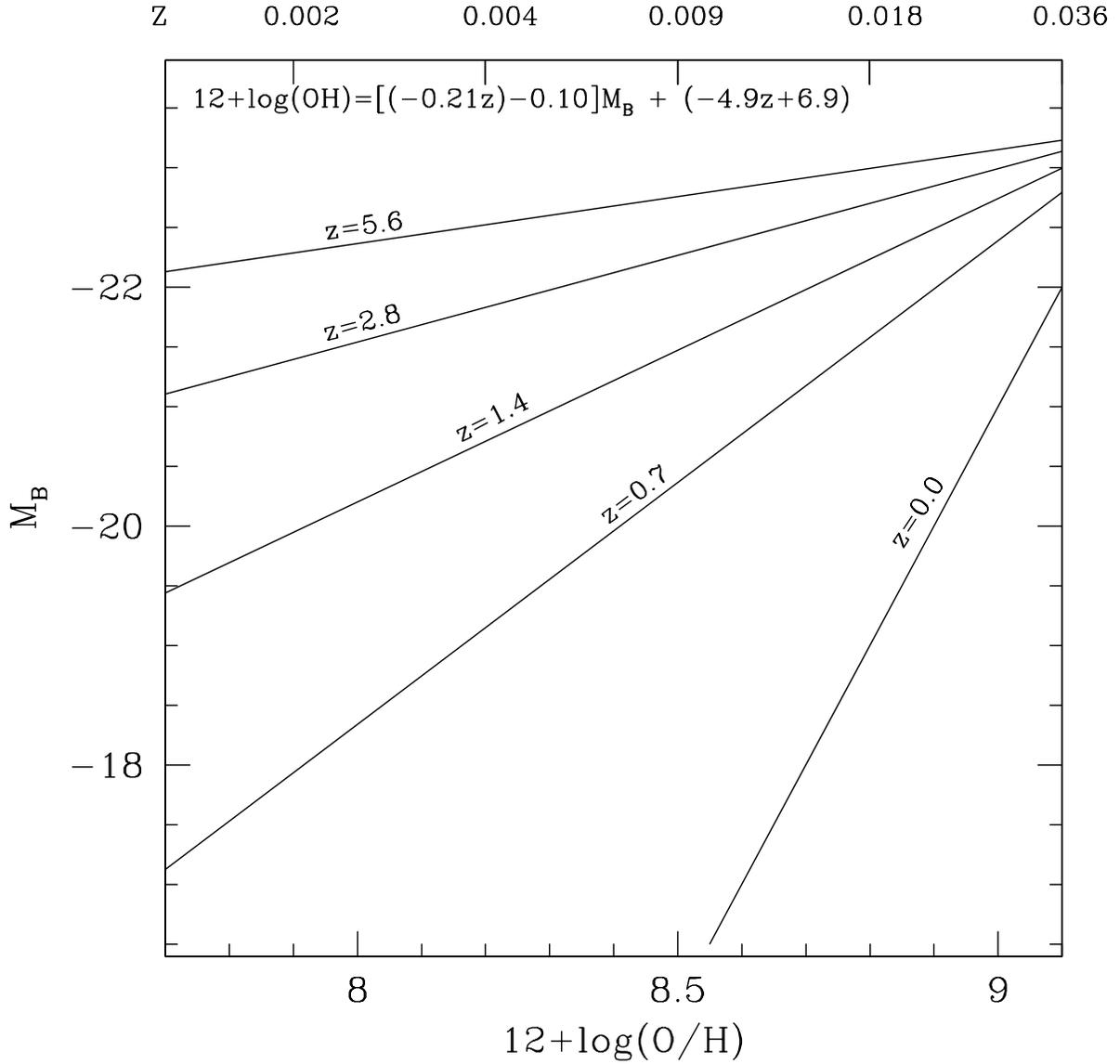}
\figcaption[LZevt.cps] 
{Hypothetical L-Z relation as a function of redshift
based upon the data in Figure~\ref{LZevp}.  
Data at z=0.7 are the most distant reliable data, and
the z=3 galaxies are highly uncertain, so this relation
should not be blindly extrapolated to high redshifts or
extremely low luminosities.
The evolution of the L-Z relation with cosmic epoch is 
plausibly due to a longer gas accretion timescale and
lower star formation rates for the lower mass galaxies.
An analytical expression approximating the
evolution of the L-Z relation is given in the figure. 
\label{LZevt} } \end{figure}

\clearpage

\begin{deluxetable}{rrrrcccccccrrrrrrrl}
\label{srctable.tab}
\rotate
\tabletypesize{\scriptsize} 
\setlength{\tabcolsep}{0.02in} 
\tablewidth{8.3in}
\tablecaption{DGSS Selected Galaxies}
\tablehead{
\colhead{ID } & 
\colhead{Object} & 
\colhead{RA (2000)}   & 
\colhead{DEC (2000)}   & 
\colhead{z}   & 
\colhead{$I_{814} AB$} & 
\colhead{$M_B$} & 
\colhead{$(B-V)$} &
\colhead{$R_{hl}$} &
\colhead{$B/T$}  &
\colhead{$R_T$}   &
\colhead{$R_A$}   &
\colhead{$EW_{3727}$}   &
\colhead{$EW_{4861}$}   &
\colhead{$EW_{4959}$}     &
\colhead{$EW_{5007}$}  &
\colhead{$12+log(O/H)$}  &
\colhead{$SFR$}  &
\colhead{Notes}  \\ [.2ex]
\colhead{(1)} &  
\colhead{(2)} &  
\colhead{(3)} &  
\colhead{(4)} &  
\colhead{(5)} &
\colhead{(6)} &
\colhead{(7)} &
\colhead{(8)} &
\colhead{(9)} &
\colhead{(10)} &
\colhead{(11)} &
\colhead{(12)} &
\colhead{(13)} &
\colhead{(14)} &
\colhead{(15)} &
\colhead{(16)}  &
\colhead{(17)}  &
\colhead{(18)}  &
\colhead{(19)}   } 
\startdata
 1 &  072-2273 &  14:17:38.928 & 52:29:34.72 & 0.2888 &  22.03 & -18.0 &  0.36 &  2.36 &  0.00 &  0.03 &  0.05 &  49.5$\pm$1.4 &  10.1$\pm$0.4 &   6.5$\pm$0.4 &  25.7$\pm$0.4 & 8.52$\pm$0.03 &   0.3 & A \\
 2 &  072-4040 &  14:17:37.838 & 52:28:58.44 & 0.2908 &  20.01 & -19.6 &  0.70 &  9.42 &  0.00 &  0.15 &  0.06 &  26.9$\pm$1.1 &   5.9$\pm$0.3 &   1.7$\pm$0.2 &   5.7$\pm$0.2 & 8.71$\pm$0.03 &   1.0 & A \\
 3 &  172-1242 &  14:16:35.570 & 52:17:27.91 & 0.3288 &  24.22 & -16.4 &  0.44 &  1.37 &  0.10 &  -.02 &  0.01 &  22.9$\pm$1.7 &  30.2$\pm$3.6 &  39.1$\pm$3.9 &    121$\pm$4  & 8.73$\pm$0.04 &   0.1 & A,B \\
 4 &  092-1375 &  14:17:26.699 & 52:27:19.48 & 0.3315 &  23.84 & -16.6 &  0.39 &  1.54 &  0.15 &  0.06 &  0.04 &  22.8$\pm$1.2 &  12.6$\pm$0.7 &  13.2$\pm$0.7 &  32.5$\pm$0.7 & 8.75$\pm$0.02 &   0.1 & A,B \\
 5 &  134-2322 &  14:17:05.547 & 52:21:18.17 & 0.3319 &  21.28 & -18.3 &  1.02 &  2.39 &  0.34 &  0.05 &  0.15 &  39.1$\pm$1.0 &   9.9$\pm$0.5 &  10.8$\pm$0.6 &  24.1$\pm$0.5 & 8.58$\pm$0.03 &   0.7 & A \\
 6 &  142-4644 &  14:16:51.341 & 52:20:52.38 & 0.3322 &  21.52 & -18.8 &  0.47 &  3.62 &  0.00 &  0.03 &  0.01 &  76.1$\pm$1.2 &  19.2$\pm$1.6 &   6.9$\pm$1.9 &  30.5$\pm$2.1 & 8.63$\pm$0.07 &   1.2 & A \\
 7 &  223-2132 &  14:16:07.432 & 52:11:25.04 & 0.3359 &  23.16 & -17.2 &  0.48 &  1.16 &  0.04 &  -.01 &  0.00 &  33.8$\pm$2.8 &   6.8$\pm$0.9 &   5.1$\pm$0.8 &  15.1$\pm$1.1 & 8.57$\pm$0.10 &   0.1 & A \\
 8 &  124-3864 &  14:17:14.790 & 52:21:50.38 & 0.3564 &  21.55 & -18.8 &  0.56 &  3.02 &  0.00 &  0.05 &  0.07 &  34.2$\pm$3.3 &   7.2$\pm$0.5 &   0.8$\pm$0.7 &   2.5$\pm$0.5 & 8.71$\pm$0.07 &   0.5 & A \\
 9 &  103-6061 &  14:17:27.434 & 52:26:08.93 & 0.3631 &  21.88 & -18.6 &  0.51 &  0.93 &  0.49 &  0.03 &  0.01 &  38.9$\pm$0.7 &  10.9$\pm$0.6 &   4.4$\pm$0.5 &  11.3$\pm$0.5 & 8.74$\pm$0.03 &   0.6 & A \\
10 &  203-4851 &  14:16:21.680 & 52:14:14.78 & 0.3639 &  22.58 & -18.1 &  0.36 &  1.18 &  0.70 &  0.02 &  0.01 &  52.6$\pm$1.1 &  12.2$\pm$0.4 &  13.8$\pm$0.4 &  34.6$\pm$0.4 & 8.42$\pm$0.02 &   0.3 & A \\
11 &  313-3557 &  14:15:11.917 & 52:01:10.88 & 0.3757 &  20.56 & -20.2 &  0.42 &  2.35 &  0.25 &  0.06 &  0.04 &  44.4$\pm$0.9 &  13.3$\pm$0.2 &   3.3$\pm$0.1 &   9.0$\pm$0.2 & 8.78$\pm$0.01 &   2.8 & A \\
12 &  304-4028 &  14:15:17.787 & 52:01:23.63 & 0.3760 &  22.78 & -17.9 &  0.41 &  1.38 &  0.02 &  0.03 &  0.04 &  99.5$\pm$1.1 &  27.8$\pm$1.3 &  14.6$\pm$1.6 &    46$\pm$1.6 & 8.64$\pm$0.02 &   0.7 & A \\
13 &  193-5319 &  14:16:24.702 & 52:15:23.83 & 0.3929 &  20.62 & -20.1 &  0.46 &  4.45 &  0.03 &  0.06 &  0.07 &  55.5$\pm$0.4 &  12.5$\pm$0.2 &   2.3$\pm$0.2 &   7.3$\pm$0.4 & 8.69$\pm$0.01 &   2.8 & A \\
14 &  304-3354 &  14:15:17.610 & 52:00:56.86 & 0.4263 &  20.57 & -20.6 &  0.29 &  1.90 &  0.51 &  0.09 &  0.12 &     113$\pm$2 &  44.3$\pm$2.4 &  30.0$\pm$2.6 &  97.1$\pm$2.6 & 8.68$\pm$0.02 &  12.7 & A \\
15 &  223-6341 &  14:16:07.359 & 52:12:07.83 & 0.4263 &  21.94 & -19.2 &  0.29 &  2.71 &  0.06 &  0.06 &  0.07 &  82.1$\pm$2.2 &    19$\pm$2.2 &  14.4$\pm$1.1 &  48.1$\pm$1.1 & 8.52$\pm$0.07 &   1.5 & A \\
16 &  292-3870 &  14:15:14.882 & 52:03:47.19 & 0.4266 &  20.90 & -20.2 &  0.36 &  2.98 &  0.00 &  0.19 &  0.08 &  63.6$\pm$2.2 &  19.9$\pm$2.2 &  11.9$\pm$1.1 &  40.0$\pm$1.1 & 8.66$\pm$0.05 &   4.2 & A \\
17 &  093-6667 &  14:17:34.454 & 52:27:25.05 & 0.4306 &  23.08 & -18.0 &  0.34 &  3.07 &  0.01 &  -.00 &  0.01 &  39.5$\pm$0.9 &  17.1$\pm$0.5 &  28.2$\pm$0.5 &  57.1$\pm$0.5 & 8.61$\pm$0.01 &   0.5 & A \\
18 &  164-3515 &  14:16:47.287 & 52:17:57.62 & 0.4320 &  21.74 & -19.3 &  0.42 &  3.23 &  0.01 &  0.02 &  0.02 &  54.8$\pm$0.8 &  12.2$\pm$1.1 &   3.2$\pm$1.3 &  14.3$\pm$1.3 & 8.64$\pm$0.05 &   1.2 & A \\
19 &  164-2417 &  14:16:46.114 & 52:17:53.61 & 0.4323 &  21.30 & -19.8 &  0.38 &  3.86 &  0.08 &  0.08 &  0.08 &  78.4$\pm$0.8 &  16.2$\pm$1.7 &   8.5$\pm$2.0 &  29.2$\pm$2.0 & 8.53$\pm$0.07 &   2.4 & A \\
20 &  074-4757 &  14:17:48.165 & 52:27:48.29 & 0.4323 &  21.49 & -19.6 &  0.40 &  4.82 &  0.03 &  0.15 &  0.09 &     103$\pm$2 &  24.7$\pm$1.5 &  11.6$\pm$1.3 &  38.8$\pm$1.5 & 8.60$\pm$0.03 &   3.1 & A \\
21 &  094-1054 &  14:17:31.025 & 52:25:24.32 & 0.4331 &  19.66 & -21.3 &  0.54 &  5.37 &  0.10 &  0.09 &  0.11 &  11.5$\pm$0.6 &   4.2$\pm$0.2 &   0.6$\pm$0.4 &   2.0$\pm$0.2 & 8.93$\pm$0.03 &   2.8 & A \\
22 &  183-4770 &  14:16:36.632 & 52:16:38.91 & 0.4691 &  21.38 & -19.9 &  0.41 &  1.85 &  0.56 &  0.05 &  0.06 &     121$\pm$1 &  45.1$\pm$3.0 &  53.7$\pm$3.2 & 126$\pm$3     & 8.60$\pm$0.03 &   7.7 & A \\
23 &  223-7714 &  14:16:04.136 & 52:12:15.01 & 0.4711 &  21.72 & -19.4 &  0.53 &  2.12 &  0.31 &  0.02 &  0.04 &    32$\pm$1.3 &   7.8$\pm$0.2 &     2$\pm$0.2 &   6.0$\pm$0.3 & 8.73$\pm$0.02 &   0.9 & A \\
24 &  062-1570 &  14:17:46.321 & 52:30:43.30 & 0.4770 &  21.38 & -19.9 &  0.45 &  2.49 &  0.34 &  0.05 &  0.06 &  26.9$\pm$0.6 &   6.9$\pm$0.4 &   4.6$\pm$0.4 &  15.9$\pm$0.4 & 8.65$\pm$0.03 &   1.2 & A \\
25 &  162-5547 &  14:16:37.359 & 52:18:33.42 & 0.4776 &  21.24 & -20.0 &  0.48 &  2.03 &  0.03 &  0.06 &  0.05 &  73.2$\pm$0.8 &  22.1$\pm$0.6 &   7.3$\pm$0.6 &  24.5$\pm$0.6 & 8.73$\pm$0.01 &   4.5 & A \\
26 &  083-6536 &  14:17:37.724 & 52:28:28.13 & 0.4798 &  23.75 & -17.6 &  0.45 &  0.44 &  0.26 &  0.03 &  -.02 &  55.5$\pm$1.7 &  14.2$\pm$1.2 &  13.6$\pm$1.3 &  38.9$\pm$1.7 & 8.55$\pm$0.05 &   0.2 & A \\
27 &  212-2260 &  14:16:08.191 & 52:13:03.10 & 0.4832 &  21.43 & -19.9 &  0.42 &  2.77 &  0.14 &  0.08 &  0.08 &  94.1$\pm$0.6 &  22.7$\pm$0.2 &  11.7$\pm$0.2 &  38.9$\pm$0.2 & 8.59$\pm$0.01 &   3.9 & A \\
28 &  303-1256 &  14:15:18.750 & 52:01:58.57 & 0.4850 &  22.09 & -19.3 &  0.41 &  5.05 &  0.52 &  0.05 &  0.08 &  98.5$\pm$4.3 &  30.9$\pm$2.5 &  16.8$\pm$2.8 &  65.3$\pm$2.8 & 8.64$\pm$0.04 &   3.0 & A \\
29 &  303-3546 &  14:15:17.146 & 52:02:18.14 & 0.4850 &  20.52 & -20.7 &  0.51 &  4.96 &  0.10 &  0.04 &  0.06 &  41.6$\pm$0.4 &  11.2$\pm$0.4 &   1.6$\pm$0.5 &   5.1$\pm$0.5 & 8.77$\pm$0.02 &   4.6 & A \\
30 &  262-5149 &  14:15:33.122 & 52:06:56.39 & 0.5084 &  23.14 & -18.3 &  0.46 &  2.64 &  0.08 &  0.02 &  -.03 &  70.6$\pm$1.0 &  21.0$\pm$2.3 &  25.5$\pm$1.9 &    92$\pm$1.9 & 8.47$\pm$0.06 &   0.8 & A \\
31 &  114-3114 &  14:17:19.113 & 52:23:47.26 & 0.5470 &  21.59 & -20.1 &  0.46 &  3.43 &  0.03 &  0.03 &  0.04 &  78.8$\pm$2.0 &  15.3$\pm$1.0 &   3.1$\pm$1.0 &   6.1$\pm$1.0 & 8.60$\pm$0.04 &   3.1 & A \\
32 &  283-3961 &  14:15:31.425 & 52:04:46.65 & 0.5586 &  23.57 & -18.1 &  0.51 &  1.07 &  0.63 &  0.04 &  0.03 &   134$\pm$2.2 &  52.3$\pm$8.6 & 71.2$\pm$10.4 &    239$\pm$11 & 8.49$\pm$0.09 &   1.8 & A \\
33 &  082-1064 &  14:17:33.800 & 52:28:19.15 & 0.6039 &  23.26 & -18.7 &  0.41 &  1.36 &  0.18 &  -.01 &  0.06 &  40.6$\pm$0.6 &   8.2$\pm$0.8 &   5.5$\pm$0.8 &  21.1$\pm$0.8 & 8.53$\pm$0.07 &   0.4 & A \\
34 &  282-2474 &  14:15:22.654 & 52:05:04.57 & 0.6315 &  21.52 & -20.6 &  0.46 &  2.35 &  0.30 &  0.03 &  0.05 &  38.2$\pm$0.7 &  12.2$\pm$0.6 &   4.7$\pm$0.3 &  10.9$\pm$0.3 & 8.78$\pm$0.02 &   3.6 & A \\
35 &  282-3252 &  14:15:22.320 & 52:04:41.62 & 0.6317 &  22.33 & -19.8 &  0.34 &  3.09 &  0.01 &  0.08 &  0.07 &     118$\pm$1 &  36.2$\pm$0.9 &  50.1$\pm$0.5 &  141$\pm$2    & 8.47$\pm$0.01 &   4.7 & A \\
36 &  212-6648 &  14:16:03.907 & 52:12:41.35 & 0.6371 &  23.66 & -18.5 &  0.36 &  1.92 &  0.02 &  0.05 &  0.09 &     117$\pm$2 &  75.1$\pm$9.7 &     152$\pm$4 &  560$\pm$10   & 8.50$\pm$0.07 &   3.1 & A \\
37 &  093-6526 &  14:17:30.101 & 52:27:15.58 & 0.6431 &  23.45 & -18.7 &  0.38 &  1.67 &  0.37 &  0.01 &  0.04 &  63.1$\pm$0.4 &  15.3$\pm$1.5 &  15.8$\pm$0.6 &  50.3$\pm$0.6 & 8.49$\pm$0.06 &   0.7 & A \\
38 &  063-5323 &  14:17:49.728 & 52:30:31.88 & 0.6437 &  21.27 & -20.9 &  0.32 &  4.29 &  0.00 &  0.19 &  0.14 &  57.0$\pm$0.2 &  19.8$\pm$0.3 &  10.8$\pm$0.3 &  34.1$\pm$0.3 & 8.72$\pm$0.01 &   7.1 & A \\
39 &  262-3751 &  14:15:34.530 & 52:07:00.73 & 0.6497 &  20.74 & -21.3 &  0.50 &  4.68 &  0.10 &  0.18 &  0.14 &  17.4$\pm$1.5 &   5.2$\pm$0.5 &   0.5$\pm$0.4 &   4.1$\pm$0.4 & 8.85$\pm$0.06 &   3.7 & A \\
40 &  284-3046 &  14:15:30.071 & 52:03:24.75 & 0.6506 &  23.35 & -18.8 &  0.46 &  1.08 &  0.77 &  0.00 &  0.02 &  79.3$\pm$1.9 &  19.8$\pm$1.9 &  23.6$\pm$2.9 &  61.4$\pm$2.9 & 8.49$\pm$0.06 &   1.1 & A \\
41 &  172-5435 &  14:16:31.329 & 52:17:11.67 & 0.6605 &  22.55 & -19.6 &  0.48 &  3.61 &  0.01 &  0.08 &  0.03 &  49.1$\pm$0.9 &  10.3$\pm$0.9 &   4.4$\pm$2.1 &  11.8$\pm$1.9 & 8.62$\pm$0.06 &   1.3 & A \\
42 &  152-1633 &  14:16:48.315 & 52:19:38.62 & 0.6736 &  21.80 & -20.5 &  0.44 &  6.99 &  0.02 &  0.06 &  0.09 & 119$\pm$2     &  27.0$\pm$0.9 &  14.3$\pm$5.7 &    44$\pm$5.7 & 8.56$\pm$0.03 &   7.2 & A \\
43 &  063-7209 &  14:17:47.868 & 52:30:47.38 & 0.6760 &  21.01 & -21.3 &  0.33 &  3.73 &  0.49 &  0.09 &  0.17 &  69.3$\pm$0.3 &  34.1$\pm$0.3 &  16.1$\pm$0.9 &  50.8$\pm$1.7 & 8.80$\pm$0.01 &  17.2 & A \\
44 &  164-3859 &  14:16:48.501 & 52:17:15.55 & 0.6828 &  22.42 & -19.9 &  0.47 &  4.62 &  0.02 &  0.03 &  0.05 &  64.1$\pm$1.9 &  14.5$\pm$1.1 &   3.9$\pm$1.9 &  11.5$\pm$1.9 & 8.65$\pm$0.05 &   2.3 & A \\
45 &  292-4369 &  14:15:14.367 & 52:03:45.35 & 0.6926 &  20.61 & -21.7 &  0.47 &  3.41 &  0.58 &  0.07 &  0.09 &  39.3$\pm$0.5 &  22.2$\pm$1.7 &  11.5$\pm$3.7 &  34.6$\pm$3.7 & 8.83$\pm$0.03 &  19.6 & A \\
46 &  282-6050 &  14:15:19.468 & 52:04:32.81 & 0.6946 &  22.70 & -19.7 &  0.39 &  2.50 &  0.01 &  0.08 &  0.12 &  97.9$\pm$1.8 &  35.4$\pm$1.8 &  29.4$\pm$2.8 &   103$\pm$3.0 & 8.61$\pm$0.03 &   4.3 & A \\
47 &  153-3721 &  14:16:51.198 & 52:19:47.44 & 0.7043 &  21.07 & -21.4 &  0.32 &  3.19 &  0.25 &  0.10 &  0.08 &  71.5$\pm$0.9 &  32.0$\pm$0.9 &  23.4$\pm$0.9 &    71$\pm$0.9 & 8.72$\pm$0.01 &  16.9 & A \\
48 &  292-7343 &  14:15:11.836 & 52:03:14.00 & 0.7450 &  23.56 & -19.0 &  0.29 &  1.64 &  0.02 &  0.03 &  0.07 &  99.1$\pm$1.8 &  24.1$\pm$4.4 &  26.3$\pm$4.5 &  79.6$\pm$3.6 & 8.46$\pm$0.12 &   1.4 & A \\
49 &  084-6809 &  14:17:42.554 & 52:27:29.20 & 0.7467 &  21.92 & -20.6 &  0.46 &  6.54 &  0.09 &  0.10 &  0.12 &     121$\pm$2 &  26.2$\pm$1.8 &  16.0$\pm$2.0 &    48$\pm$2.0 & 8.52$\pm$0.04 &   8.0 & A \\
50 &  182-7536 &  14:16:22.681 & 52:15:59.33 & 0.7505 &  22.73 & -19.8 &  0.40 &  3.86 &  0.04 &  0.07 &  0.08 &  72.5$\pm$1.8 &  18.2$\pm$1.2 &  13.7$\pm$1.4 &  37.9$\pm$1.4 & 8.57$\pm$0.04 &   2.4 & A \\
51 &  292-6724 &  14:15:12.920 & 52:02:56.62 & 0.7659 &  21.93 & -20.7 &  0.47 &  3.37 &  0.09 &  0.06 &  0.10 &  35.3$\pm$0.9 &  11.3$\pm$0.9 &   4.2$\pm$1.6 &  11.8$\pm$1.6 & 8.77$\pm$0.04 &   3.7 & A \\
52 &  134-0967 &  14:17:05.035 & 52:20:31.42 & 0.7885 &  22.83 & -19.9 &  0.26 &  3.76 &  0.00 &  0.14 &  0.07 &     117$\pm$2 &  51.9$\pm$1.9 &  58.1$\pm$2.5 &  225$\pm$3    & 8.55$\pm$0.02 &   6.9 & A \\
53 &  142-4838 &  14:16:51.266 & 52:20:45.96 & 0.8077 &  20.18 & -22.6 &  0.28 &  5.17 &  0.25 &  0.16 &  0.09 &  13.1$\pm$1.9 &  11.4$\pm$0.5 &   9.5$\pm$0.7 &  33.3$\pm$0.5 & 8.80$\pm$0.02 &  18.5 & A \\
54 &  174-3527 &  14:16:40.994 & 52:16:35.72 & 0.8096 &  21.59 & -21.2 &  0.46 &  3.19 &  0.00 &  0.10 &  0.18 &  56.9$\pm$1.7 &  25.8$\pm$3.1 &  10.4$\pm$2.1 &  23.9$\pm$2.1 & 8.84$\pm$0.04 &  13.2 & A \\
55 &  152-3226 &  14:16:46.841 & 52:19:28.70 & 0.8127 &  22.52 & -20.3 &  0.34 &  4.42 &  0.42 &  0.05 &  0.10 &     109$\pm$2 &  28.3$\pm$4.4 &  22.9$\pm$3.0 &  88.8$\pm$3.0 & 8.50$\pm$0.09 &   5.4 & A \\
56 &  153-6078 &  14:16:56.673 & 52:20:22.44 & 0.8128 &  20.98 & -21.8 &  0.48 &  6.87 &  0.06 &  0.14 &  0.15 &  70.0$\pm$5.2 &  59.5$\pm$7.9 &   7.7$\pm$1.5 &  23.4$\pm$1.5 & 9.00$\pm$0.02 &  58.3 & A \\
57 &  073-2658 &  14:17:47.597 & 52:29:03.36 & 0.2863 &  21.13 & -18.9 &  0.34 &  2.88 &  0.28 &  0.04 &  0.10 &  62.1$\pm$1.6 &  10.8$\pm$1.1 &  11.1$\pm$2.2 &  23.0$\pm$1.2 & 8.44$\pm$0.07 &   0.7 & C \\
58 &  173-5210 &  14:16:36.664 & 52:17:39.89 & 0.3570 &  20.03 & -20.4 &  0.54 &  4.51 &  0.05 &  0.07 &  0.06 &  36.9$\pm$0.9 &   5.8$\pm$0.4 &   1.4$\pm$0.4 &   3.5$\pm$0.4 & 8.58$\pm$0.05 &   1.7 & C \\
59 &  213-6640 &  14:16:13.664 & 52:13:20.44 & 0.3656 &  20.22 & -20.3 &  0.50 &  4.62 &  0.03 &  0.06 &  0.05 &  29.2$\pm$1.4 &   3.5$\pm$0.3 &   0.5$\pm$0.3 &   0.8$\pm$0.3 & 8.51$\pm$0.10 &   0.9 & C \\
60 &  183-6415 &  14:16:30.453 & 52:16:42.93 & 0.3865 &  21.00 & -19.6 &  0.55 &  7.06 &  0.00 &  0.14 &  0.07 &     131$\pm$4 &  24.2$\pm$2.5 &   8.3$\pm$2.5 &  27.8$\pm$2.7 & 8.51$\pm$0.07 &   3.6 & C \\
61 &  092-1962 &  14:17:26.421 & 52:27:06.04 & 0.4261 &  19.49 & -21.5 &  0.40 &  6.35 &  0.03 &  0.20 &  0.15 &  25.7$\pm$0.2 &   4.3$\pm$0.2 &   1.0$\pm$0.1 &   4.3$\pm$0.1 & 8.64$\pm$0.03 &   3.3 & C \\
62 &  313-7545 &  14:15:09.713 & 52:01:46.44 & 0.4503 &  22.70 & -18.4 &  0.42 &  2.54 &  0.01 &  0.01 &  0.01 & 54.9$\pm$12.1 &   9.1$\pm$0.8 &   6.8$\pm$1.1 &  24.6$\pm$1.1 & 8.42$\pm$0.14 &   0.4 & C \\
63 &  183-1153 &  14:16:35.713 & 52:16:00.42 & 0.5070 &  20.74 & -20.8 &  0.39 &  5.96 &  0.42 &  0.09 &  0.13 &  64.3$\pm$2.1 &  10.6$\pm$0.6 &   3.6$\pm$0.6 &  15.5$\pm$0.6 & 8.48$\pm$0.05 &   3.9 & C \\
64 &  293-4412 &  14:15:19.779 & 52:03:30.50 & 0.6462 &  22.90 & -19.3 &  0.30 &  6.21 &  0.00 &  0.04 &  0.05 &     175$\pm$2 &  32.8$\pm$2.1 &  20.2$\pm$1.9 &  96.5$\pm$1.9 & 8.38$\pm$0.04 &   2.8 & C \\
65 &  092-7832 &  14:17:20.880 & 52:26:23.36 & 0.6824 &  23.57 & -18.7 &  0.29 &  1.43 &  0.54 &  0.00 &  0.10 &  59.0$\pm$1.9 &   154$\pm$0.7 &   117$\pm$4   &   237$\pm$4   & 8.93$\pm$0.04 &   8.2 & D \\
66 &  203-3109 &  14:16:17.618 & 52:13:49.43 & 0.6848 &  22.41 & -19.9 &  0.34 &  5.60 &  0.02 &  0.10 &  0.08 &  60.5$\pm$1.7 &  30.2$\pm$1.5 &  11.3$\pm$2.2 &  31.1$\pm$2.2 & 8.85$\pm$0.02 &   4.9 & D \\
\enddata
\tablerefs{
(1) Reference ID \# for this paper;
(2) Groth Strip survey ID ; 
(3) J2000 Right Ascension; 
(4) J2000 Declination; 
(5) redshift from optical emission lines
(6) $I_{814}$ AB magnitude
(7) Rest-frame $M_B$ for $H_0=70$ km/s/Mpc, $\Omega_M=0.3$, $\Omega_\Lambda=0.7$
(8) Rest-frame B-V color
(9) Half light radius in kpc derived from the global $I_{814}$ image
(10) Bulge fraction derived from the $I_{814}$ image.
(11) $R_T$ asymmetry index as defined in paper II
(12) $R_A$ asymmetry index as defined in Paper II
(13) EW of [O~II] $\lambda$3727 and uncertainty corrected to the rest frame
(14) EW of H$\beta$ $\lambda$4861 and uncertainty corrected to the rest frame
(15) EW of [O~III] $\lambda$4959 and uncertainty corrected to the rest frame
(16) EW of [O~II] $\lambda$5007 and uncertainty corrected to the rest frame
(17) Oxygen abundance, 12+log(O/H), from the empirical $R_{23}$ method
following McGaugh (1991) as formulated in KKP after correction for \AA\ of 
stellar absorption in the $EW_{H\beta}$.  An additional uncertainty of 
$\sim$0.15 dex in O/H representing uncertainties in the
photoionization models and empirical strong-line
calibration should be added in quadrature to the tabulated
measurement errors. ;
(18) Estimated star formation rate based on $H\beta$ flux 
(not corrected for stellar absorption), derived from the
	EW($H\beta$) and V-band absolute magnitude.  We estimate the
	$H\beta$ luminosity as $L_{H\beta}(erg/s)=5.49\times10^{31} \times 2.5^{-M_V} \times EW_{H\beta}$.
	The SFR then is then computed by
	$SFR(M_\odot/yr)= 2.8 \times L_{H\beta} /1.12\times10^{41}$
	which assumes the Kennicutt (1983) calibration of SFR in terms
	of $H\alpha$ luminosity.  This estimate is a lower limit since extinction
	and stellar Balmer absorption is not taken into account here.  
(19) Notes: 
A--Probable star-forming galaxy (non-AGN or minimal AGN contribution), as defined in text;
B--Low-luminosity object with line ratios in the
``turn-around'' region of the strong-line abundance
diagram, so the oxygen abundance is 
highly uncertain, in the range $7.8<12+log(O/H)<8.4$.  Object removed
from remaining analysis.
C--Object is probable AGN on basis of Figure~\ref{AGNtest}; 
D--Object is a probable AGN on basis of [Ne III]/[O~II] $>0.4$}
\end{deluxetable}


\begin{references}

\reference{} Anders, E., \& Grevesse, N. 1989, GeCoA, 53, 197
\reference{} Baldry, I.~K. \etal\ 2002, ApJ, 569, 582
\reference{} Baldwin, J.~A., Phillips, M.~M., \& Terlevich, R. 1981, PASP,
	93, 5
\reference{} Bergeron, J., \& Stasinska, G. 1986, A\&A, 169, 1
\reference{} Bershady, M.~A., Haynes, M.~P., Giovanelli, R., \&
Andersen, D.~R. 1988, in Galaxy Dynamics, eds. D.~R. Merritt, M.
Valluri, \& J.~A. Sellwood (ASP Conf Series)
\reference{} Blumenthal, G.~M., Faber, S.~M., Primack, J.~R., \& Rees,
M.~J. 1984, Nature, 311, 517
\reference{} Brodie, J.~P., \& Huchra, J.~P. 1991, ApJ, 379, 157
\reference{} Calzetti, D., Kinney, A.~L., \& Storchi-Bergmann, T. 1994, 
	ApJ, 429, 582 
\reference{} Carollo, C., M. \& Lilly, S.~J. 2001, ApJ, 548, L153 (CL01)
\reference{} Cohen, J.~G. 2002, ApJ, 567, 672
\reference{} De Young, D.~S., \& Gallagher, J.~S. III 1990, ApJ, 356, L15 
\reference{} Djorgovski, S., \& Davis, M. 1987, ApJ, 313, 59
\reference{} Dopita, M.~A., \& Evans, I.~N. 1986, ApJ, 307, 431
\reference{} Dressler, A., Lynden-Bell, D., Burstein, D., Davies,
	R.~L., Faber, S.~M., Terlevich, R.~J., \& Wegner, G. 1987, ApJ, 
	313, 42
\reference{} Edmunds, M.~G. \& Pagel, B.~E.~J. 1984, MNRAS, 211, 507
\reference{} Faber, S.~M. 1973, ApJ, 179, 423
\reference{} Fioc, M. \& Rocca-$\!$Volmerange, B. 1999, astro-ph/9912179 
\reference{} Forbes, D.~A., Phillips, A.~C., Koo, D.~C., \& Illingworth, 
	G.~D. 1995, ApJ, 462, 89
\reference{} French, H.~B., 1980, ApJ, 240, 41
\reference{} Garnett, D.~R. 2002, ApJ, 581, 1019
\reference{} Gebhardt, K., Faber, S.~M., Koo, D.~C., Im, M., Simard, L.,
    Illingworth, G.~D., Phillips, A.~C., Sarajedini, V.~L., Vogt, N.~P., 
    \& Willmer, C.~N.~A. 2003, ApJ, in press
\reference{} Groth, E. J., Kristian, J.~A., Lynds, R., O'Neil, E.~J. Jr.,
	Balsano, R., \& Rhodes, J. 1994, BAAS, 185 
\reference{} Heckman, T.~M. 1980, A\&A, 87, 152
\reference{} Heckman, T.~M., Lehnert, M~D., Strickland, D.~K.,
	\& Armus, L. 2000, ApJS, 129, 493
\reference{} Huchra, J.~P., Davis, M., Latham, D., \& Tonry, J. 1983, 
	ApJS, 52, 89
\reference{} Im, M., Simard, L., Faber, S.~M., Koo, D.~C., Gebhardt, K.,
	Willmer, C.~N.~A., Phillips, A.~C., Illingworth, G.~D., Vogt, N.~P., 
	\& Sarajedini, V.~L. 2002, ApJ, 571, 136 
\reference{} Jansen, R.~A., Franx, M., Fabricant, D., \& Caldwell, N.
	2000a, ApJS, 126, 271 (NFGS)
\reference{} Jansen, R.~A., Fabricant, D., Franx, M., \& Caldwell, N.
	2000b, ApJS, 126, 331
\reference{} Kannappan, S.~J., Fabricant, D.~G., \& Franx, M. 2002, AJ,
	123, 2358
\reference{} Kauffman, G. 1996, MNRAS, 281, 475
\reference{} Kelson, D.~D., van Dokkum, P.~G., Franx, M., Illingworth,
	G., \& Fabricant, D. 1997, ApJ, 478, L13
\reference{} Kennicutt, R.~C. Jr. 1983, ApJ, 272, 54
\reference{} Kennicutt, R.~C. Jr. 1992a, ApJS, 79, 255 
\reference{} Kennicutt, R.~C. Jr. 1992b, ApJ, 388, 310 
\reference{} Kennicutt, R.~C. Jr. 1998, ApJ, 498, 541 
\reference{} Kennicutt, R.~C. Jr. \& Skillman, E.~D.
	2001, AJ, 121, 1461
\reference {} Kinney, A. L., Calzetti, D., Bohlin, R. C., McQuade, K.,
	Storchi-Bergmann, T., \& Schmitt, H. R.  1996, ApJ, 467, 38
\reference{} Kobulnicky, H.~A., Kennicutt, R.~C., \& Pizagno, J. 1998,
        ApJ, 514, 544
\reference{} Kobulnicky, H.~A. \& Koo, D.~C. 2000, ApJ, 545, 712 (KK00)
\reference{} Kobulnicky, H.~A. \& Phillips, A.~C. 2003, ApJ, 000 (KP03)
\reference{} Kobulnicky, H.~A. \& Zaritsky, D. 1999, ApJ, 511, 118 (KZ99)
\reference{} Lequeux, J., Peimbert, M., Rayo, J.~F., Serrano, A., \& 
   Torres--Peimbert, S. 1979, A\&A, 80, 155
\reference{} Lin, H., Yee, H.~K., Carlberg, R.~G., Morris, S.~L.,
	Sawicki, M., Patton, D.~R., Wirth, G., \& Shepard, C.~W.
	1999, ApJ, 518, 533
\reference{} Lilly, S.~J., Le F\'evre, O., Hammer, F., \& Crampton, D. 1996, 
	ApJ, 460, L1
\reference{} Lilly, S.~J., Tresse, L., Hammer, F., Crampton, D.,
	\& Le F\'evre, O. 1995, ApJ, 455, 108
\reference{} Lilly, S.~J., \etal\ 1998, ApJ, 500, 75
\reference{} Lu, L., Sargent, W.~L.~W., Barlow, T.~.A., 1997, ApJ, 484, 131
\reference{} MacLow, M.~M, \& Ferrara, A.  1999, ApJ, 513, 142
\reference{} Madau, P., Ferguson, H., Dickinson, M., Giavalisco, M., 
	Steidel, C.,\& Fruchter, A. 1996, MNRAS, 283, 1388
\reference{} Marzke, R.~O., da Costa, L.~N., Pellegrini, P.~S., 
	Willmer, C.~N.~A., Geller, M.~J. 1998, ApJ, 503, 617 
\reference{} Martin, C.~L., Kobulnicky, H.~A., \& Heckman, T.~M. 2002,
	ApJ, 574, 663
\reference{} McCall, M.~L., Rybski, P.~M., \& Shields, G.~A. 1985, 
	ApJS, 57, 1 (MRS)
\reference{} McGaugh, S. 1991, ApJ, 380, 140
\reference{} McGaugh, S. 1998, private communication
\reference{} Mehlert, D. \etal\ 2002, A\&A, 393, 809
\reference{} Melbourne, J., \& Salzer, J.~J. 2002, AJ, 123, 2302
\reference{} Navarro, J.~P., Frenk, C.~S., \& White, S.~D.~M. 1995,
	MNRAS, 275, 561
\reference{} Norberg, P., \etal\ (The 2dFGRS Team), 2002, MNRAS, 336, 907
\reference{} Oke, J. B., \etal\ 1995, PASP, 107, 375
\reference{} Osterbrock, D.~E. 1989,  Astrophysics of Gaseous Nebulae
   and Active Galactic Nuclei, University Science Books:Mill Valley CA
\reference{} Pagel, B.~E.~J. Edmunds, M.~G., Blackwell, D.~E., Chun, 
		M.~S., \& Smith, G. 1979, MNRAS, 189, 95
\reference{} Pagel, B.~E.~J. 1997, ``Nucleosynthesis and Chemical
	Evolution of Galaxies'', Cambridge University Press
\reference{} Pei, Y.~C., \& Fall, S.~M. 1995, ApJ, 454, 69
\reference{} Pettini, M., Shapley, A.~E., Steidel, C.~C., Cuby, J.-G.,
	Dickinson, M., Moorwood, A.~F.~M., Adelberger, K.~L., \& Giavalisco, M.
	2001, ApJ, 554, 981 (Pe01)
\reference{} Pettini, M., Smith, L.~J., King, D.~L., 
	Hunstead, R.~W. 1997, APJ, 486, 665  
\reference{} Prieto, C.~A., Lambert, D.~L., \& Asplund, M. 2001, ApJ, 556 L63
\reference{} Prochaska, J.~X. \& Wolf, A.~M. 1999, ApJS, 121,369
\reference{} Richer, M.~G., \& McCall, M.~L. 1995, ApJ, 445, 642
\reference{} Rola, C.~S., Terlevich, E., \& Terlevich, R.~J. 1997, MNRAS, 289, 419
\reference{} Roche, N., Ratnatunga, K., Griffiths, R.~E., Im, M., \&
	Naim, A. 1998, MNRAS, 293, 157
\reference{} Salzer, J.~J., Gronwall, C., Lipovetsky, V.~A.; Kniazev, A.,
	 Moody, J.~W,, Boroson, T.~A., Thuan, T.~ X., Izotov, Y.~I.,
	 Herrero, J.~L., Frattare, L.~M. 2000, AJ, 120, 80 (KISS)
\reference{} Sandage, A., Freeman, K.~C., \& Stokes, N.~R. 1970, ApJ, 160, 831
\reference{} Sarajedini, V.~L., \& the DEEP Team, 2003, in prep. (Paper XIII).
\reference{} Sargent, W.~L.~W., Steidel, C.~C., \& Boksenberg, A. 1988,
	ApJS, 68, 539
\reference{} Sawicki, M.~J., Lin, H., \& Yee, H.~K. 1997, AJ, 113, 1
\reference{} Schade, D., Carlberg, R.~G., Yee, H.~K., \& Lopez-Cruz, O.
	1996a, ApJ, 464, 63
\reference{} Schade, D., Carlberg, R.~G., Yee, H.~K., \& Lopez-Cruz, O.
	1996b, ApJ, 465, 103
\reference{} Schmidt, M. 1959, ApJ, 129. 243
\reference{} Simard, L., \& Pritchet, C.~J. 1998, ApJ, 505, 96
\reference{} Simard, L., Koo, D.~C., Faber, S.~M., Sarajedini, V.~L.,
  Vogt, N.~P., Phillips, A.~C., Gebhardt, K., Illingworth, G.~D., \& Wu,
  K.~L. 1999, ApJ, 1999, 519, 563
\reference{} Simard, L., Willmer, C.~N.~A., Vogt, N.~P., Sarajedini,
	V.~L., Phillips, A.~C., Koo, D.~C., Im, M., Illingworth, G.~D.,
	Gebhardt, K., \& Faber, S.~M. 2002, ApJS, 142, 1 (Paper II)
\reference{}  Skillman, E.~D., Kennicutt, R.~C., \& Hodge, P. 1989,
	ApJ, 347, 875 
\reference{} Somerville, R.~S., \& Primack, J.~R. 1999, MNRAS, 310, 1087
\reference{} Steidel, C.~C. 1990, ApJS, 74, 37
\reference{} Trager, S. C., Worthey, G., Faber, S. M., Burstein, D., \& Gonzalez, 
	J. J. 1998, ApJS, 116, 1
\reference{} Tully, R.~B., \& Fisher, J.~R. 1977, A\&A, 54, 661
\reference{} Vader, P. 1987, ApJ, 317, 128
\reference{} van Dokkum, P.~G., \& Franx, M. 1996, MNRAS, 281, 985
  \reference{} Veilleux, S., \& Osterbrock, D.~E. 1987, ApJS, 63, 295
\reference{} Vogt, N., Forbes, D., Phillips, A.~C., Gronwall, C., Faber, S.~M.,
	Illingworth, G.~D., \& Koo, D.~C. 1996, ApJ, 465, L15
\reference{} Vogt, N., Phillips, A.~C., Faber, S.~M., Gallego, J., Gronwall, C.,
Guzman, R., Illingworth, G., Koo, D.~C., \& Lowenthal, J.~D. 1997, 479, L121  
\reference{} Vogt, N. \etal\ 2002, ApJ, in prep, (Paper I)
\reference{} Walter, D.~K., Dufour, R.~J., \& Hester, J.~J. 1992, ApJ,
	397, 196
\reference{} Woosley, S. E. \& Weaver, T. A. 1995, ApJS, 101, 181
\reference{} Zaritsky, D., Kennicutt, R.~C., \& Huchra, J.~P. 1994, ApJ, 420,
	87 
\reference{}Ziegler, B.~L., B\"ohm, A., Fricke, K.~J., J\"ager, K., Nicklas, H., 
Bender, R., Drory, N., Gabasch, A., Saglia, R.~P., Seitz, S., Heidt, J., 
Hehlert, D., M\"ollenhoff, C., Noll, S., \& Sutorius, E. 2002, ApJ, 564, 
L69
\reference{} Zucca, E. \etal\ ESO Slice Project team 1997, A\&A, 326,
	477
\end{references}
\end{document}